\documentclass[superscriptaddress,aps,prd,nofootinbib]{revtex4-2}
\pdfoutput=1

\usepackage{graphicx}
\usepackage{dcolumn}
\usepackage{hyperref}
\usepackage{ulem}
\usepackage{color}
\usepackage{bm}
\usepackage{siunitx}
\usepackage{braket}
\usepackage{amsfonts,amsmath,amssymb,bm,tensor}
\usepackage{ifpdf}
\usepackage{slashed}
\usepackage[mathscr]{eucal}
\usepackage[utf8]{inputenc}
\usepackage{cancel}
\usepackage{enumerate}
\usepackage{tikz}
\usepackage{tikz-3dplot}
\usepackage[compat=1.1.0]{tikz-feynhand}
\usepackage{subfigure}
\begin{document}


\title{Gravitational wave trispectrum in the axion-SU(2) model}
\author{Tomohiro Fujita}
\affiliation{Waseda Institute for Advanced Study, Waseda University, Shinjuku, Tokyo 169-8050, Japan}
\affiliation{Research Center for the Early Universe,  The University of Tokyo, Bunkyo, Tokyo 113-0033, Japan}
\author{Kai Murai}
\affiliation{ICRR, University of Tokyo, Kashiwa, 277-8582, Japan}
\affiliation{Kavli IPMU (WPI), UTIAS, University of Tokyo, Kashiwa, 277-8583, Japan}
\author{Ippei Obata}
\affiliation{Max-Planck-Institut f{\"u}r Astrophysik, Karl-Schwarzschild-Str. 1, 85748 Garching, Germany}
\author{Maresuke Shiraishi}
\affiliation{Department of General Education, National Institute of Technology,
Kagawa College, 355 Chokushi-cho, Takamatsu, Kagawa 761-8058, Japan}

\begin{abstract}
	We study the trispectrum of the gravitational waves (GWs) generated through the dynamics of an axionic spectator field and SU(2) gauge fields during inflation.
	In non-Abelian gauge theory, the gauge fields have four-point self-interactions, which induce the tree-level GW trispectrum.
	We formulate this type of the GW trispectrum including the non-dynamical contributions and evaluate it in the equilateral limit as a unique signal of this model.
	We find that the ratio of the GW trispectrum to the cube of the scalar power spectrum can be as large as $\mathcal{O}(10^6)$ in the viable parameter space, which could be captured in the CMB observations.
\end{abstract}

\maketitle

\tableofcontents

\section{Introduction}
\label{sec: introduction}

Inflation is regarded as a part of the standard paradigm of modern cosmology
in which the structures observed in the present universe and their seed fluctuations
are consistently explained.
Inflation generally predicts not only the density perturbation but also the primordial gravitational waves (PGWs).
The amplitude of PGWs is conventionally characterized by the tensor-to-scalar ratio $r$, which is probed through the B-mode polarization in the Cosmic Microwave Background (CMB) by a number of observations.
Planck and BICEP2/Keck Array joint observations currently set the upper bound of $r \lesssim 0.06$~\cite{Akrami:2018odb,Ade:2018gkx}.
In addition, the observational sensitivity to $r$ is expected to improve up to $\Delta r = \mathcal{O}(10^{-3})$ by future experiments such as LiteBIRD satellite~\cite{Matsumura:2013aja} and CMB-S4 project~\cite{Abazajian:2016yjj,Abazajian:2020dmr}.
In the standard inflationary scenario, PGWs originate from the vacuum fluctuations during inflation and then the value of $r$ is directly related to the energy scale of inflation.
Therefore the detection of the B-mode polarization or PGWs is extremely important to determine the inflationary energy scale and to explore viable inflationary models.

However, the detection of B-mode polarization does not necessarily mean the detection of the vacuum tensor fluctuation. This is because PGWs can be produced by other sources than the vacuum fluctuations during inflation. If the sourced PGWs dominate the total PGWs, the tensor spectrum does not necessarily follow the properties of the vacuum fluctuations such as nearly scale invariant, statistically isotropic, parity symmetric, and almost Gaussian.
Therefore, in order to interpret the detection of $r$, it is crucial to understand the possible properties of sourced PGWs that are added to the vacuum fluctuations.

There are some models generating sourced PGWs where a background motion of a scalar field amplifies perturbations of a gauge field through a kinetic or topological coupling, which violates the conformal invariance of a gauge field.
Such a model is originally suggested in~\cite{Ratra:1991bn} and revisited in~\cite{Garretson:1992vt,Field:1998hi,Giovannini:2001nh,Bamba:2003av,Anber:2006xt,Martin:2007ue,Demozzi:2009fu,Kanno:2009ei,Fujita:2012rb,Ferreira:2013sqa,Fujita:2014sna,Kobayashi:2014sga,Ferreira:2014hma,Obata:2014qba,Fujita:2015iga,Fujita:2016qab,Adshead:2016iae,Vilchinskii:2017qul,Caprini:2017vnn,Sharma:2017eps,Sobol:2018djj,Fujita:2019pmi,Shtanov:2020gjp,Talebian:2020drj,Sobol:2020lec} in the context of primordial magnetogenesis.
In these models, the amplified gauge field also enhances other fluctuations including the scalar and tensor perturbations.
Depending on the setup of the models,
the curvature perturbations sourced by the gauge field can be highly non-Gaussian~\cite{Barnaby:2010vf,Barnaby:2011qe,Barnaby:2012tk,Anber:2012du,Barnaby:2012xt,Bartolo:2012sd,Linde:2012bt,Fujita:2013qxa,Ohashi:2013qba,Ferreira:2014zia},
statistically anisotropic~\cite{Bartolo:2012sd,Ohashi:2013qba,Watanabe:2009ct,Himmetoglu:2009mk,Gumrukcuoglu:2010yc,Watanabe:2010fh,Kanno:2010nr,Watanabe:2010bu,Soda:2012zm,Naruko:2014bxa,Abolhasani:2015cve},
and large enough to form primordial black holes~\cite{Linde:2012bt,Garcia-Bellido:2016dkw,Domcke:2017fix,Garcia-Bellido:2017aan,Cheng:2018yyr,Kawasaki:2019hvt,Ozsoy:2020kat}.
Moreover, the sourced PGWs can dominate the total PGWs and do not necessarily have the same properties as the PGWs generated from the vacuum fluctuations, i.e., they can be scale-dependent, statistically anisotropic, parity-violating, and/or highly non-Gaussian~\cite{Barnaby:2011qe,Anber:2012du,Barnaby:2012xt,Ohashi:2013qba,Watanabe:2010fh,Watanabe:2010bu,Kawasaki:2019hvt,Sorbo:2011rz,Cook:2011hg,Mukohyama:2014gba,Choi:2015wva,Namba:2015gja,Ito:2016aai,Domcke:2016bkh,Guzzetti:2016mkm,Peloso:2016gqs,Obata:2016oym,Obata:2016xcr,Fujita:2017jwq,Ozsoy:2017blg,Fujita:2018zbr,Obata:2018ilf,Ozsoy:2020ccy,Ozsoy:2021onx}.
Such PGWs with non-standard properties are potentially testable with the correlations of CMB temperature and polarization anisotropies~\cite{Saito:2007kt,Shiraishi:2013kxa,Bartolo:2014hwa,Bartolo:2015dga,Shiraishi:2016yun,Shiraishi:2016mok,Bartolo:2017sbu,Thorne:2017jft,Hiramatsu:2018vfw,Fujita:2018ndp}, laser interferometers~\cite{Seto:2006hf,Seto:2006dz,Seto:2008sr,Bartolo:2018qqn,Campeti:2020xwn}, or pulsar timing arrays~\cite{Kato:2015bye}. There are already various CMB constraints extracted from two-point correlations (equivalently power spectra) \cite{Saito:2007kt,Planck:2015sxf,Gerbino:2016mqb} and three-point ones (equivalently bispectra) \cite{Shiraishi:2014ila,Planck:2015zfm,Planck:2019kim}, while no detection of the sourced signal has been reported so far. This fact encourages further investigation with less used observables such as four-point correlations (equivalently trispectra). In an inflationary model where a U(1) gauge field couples to an axionic inflaton field, in the scalar sector, the trispectra can surpass in detectability the power spectra and bispectra in specific parameter space \cite{Shiraishi:2016mok}. One might be able to expect similar results also in the tensor sector.

Among the various classes of models, the models where SU(2) gauge fields are topologically coupled to an axionic field have attracted much attention in the last decade.
This class of models was originally proposed as inflation models  in~\cite{Maleknejad:2011jw,Adshead:2012kp}.
In these models, a large coupling between axion and gauge fields leads to a non-zero vacuum expectation value of SU(2) gauge fields, which is an isotropic attractor solution~\cite{Maleknejad:2013npa,Domcke:2018rvv,Wolfson:2020fqz,Wolfson:2021fya}.
Interestingly enough, the isotropic gauge field background enables gauge field perturbations to have effective tensor components. 
These tensor perturbations of gauge fields linearly couple to tensor metric perturbations (i.e. gravitational waves) and can significantly source them.
Since only one polarization mode of tensor perturbations of gauge fields exponentially grows through a tachyonic instability around the horizon crossing and it couples to the gravitational waves with the same polarization, the generated GWs are fully chiral~\cite{Dimastrogiovanni:2012ew}.
Due to this large enhancement of GWs, however, the original scenario has been excluded from CMB data~\cite{Adshead:2013qp,Adshead:2013nka}.
This conflict can be evaded in extended models by introducing additional fields~\cite{Obata:2014loa,Obata:2016tmo,Maleknejad:2016qjz,Dimastrogiovanni:2016fuu,Adshead:2016omu,DallAgata:2018ybl}.
In these models, the enhancement of GWs is consistent with the observational constraints and can be a distinctive signal of the scenario.
In addition to the GW power spectrum, the non-Gaussianities of the sourced PGWs are also important observables in this scenario.
The bispectra of GWs~\cite{Agrawal:2017awz,Agrawal:2018mrg} and the scalar-tensor-tensor correlations~\cite{Dimastrogiovanni:2018xnn,Fujita:2018vmv} have been computed and their detectabilities have been discussed \cite{Planck:2019kim,Shiraishi:2019yux}.
On the other hand, the one-loop contribution from the tensor perturbation to the curvature power spectra constrains the parameter space of the model, because it would make the curvature perturbation too non-Gaussian to be consistent with the observations~\cite{Dimastrogiovanni:2018xnn,Papageorgiou:2018rfx,Papageorgiou:2019ecb}.
The recent studies on this scenario have almost worked out the relevant non-linear processes at the leading order.

In this paper, we focus on the model involving an axion-SU(2) gauge coupling proposed in~\cite{Dimastrogiovanni:2016fuu} and compute the trispectrum of chiral GWs for the first time. We then find that the resultant trispectrum, dubbed as $\langle \psi^4 \rangle$, has two different contributions: the so-called $\tau_{\rm NL}$ and $g_{\rm NL}$ type ones that are composed of the first-order and second-order terms of the GW, $\langle \psi^4 \rangle_{\tau_{\rm NL}} \sim \langle \psi_1^2 \psi_2^2 \rangle$, and the first-order and third-order ones, $\langle \psi^4 \rangle_{g_{\rm NL}} \sim \langle \psi_1^3 \psi_3 \rangle$, respectively.%
\footnote{This classification and naming have been conventionally utilized in the scalar-mode trispectrum analyses \cite{Mizuno:2010by,Izumi:2011di}.}
The $\tau_{\rm NL}$ and $g_{\rm NL}$ type contributions arise from the three-point and four-point vertices of tensor perturbations in the Lagrangian, respectively. In inflationary models with an axion-U(1) gauge coupling, only the $\tau_{\rm NL}$ type contribution can be generated because of the existence of the three-point interaction alone \cite{Shiraishi:2016mok}. In contrast, very interestingly, in our case, since SU(2) non-Abelian gauge fields have second-order terms in its field strength, there are not only three-point but also four-point vertices, inducing nonvanishing $g_{\rm NL}$ type contribution. Therefore, the $g_{\rm NL}$ type trispectrum will become a unique probe of the axion-SU(2) gauge coupling.

From an observational point of view, the $g_{\rm NL}$ type trispectrum is easier to deal with. Unlike the $\tau_{\rm NL}$ type, the form of the $g_{\rm NL}$ type trispectrum is free from a convolution with respect to internal momenta. This fact makes the form of the resultant CMB trispectrum simple and factorizable and enables the full shape analysis by means of fast feasible estimators \cite{Fergusson:2010gn,Sekiguchi:2013hza,Smith:2015uia}, yielding optimal constraints.

Motivated by these advantages, this paper investigates the $g_{\rm NL}$ type signal thoroughly.
To evaluate the magnitude and observability of the $g_{\rm NL}$-type trispectra, we expand the Lagrangian up to the fourth order.
Furthermore, we consider the contribution of the non-dynamical components of the gauge field perturbations, which affect only the fourth or higher order terms.
Calculating the leading contributions from tree-level diagrams, 
we estimate the $g_\mathrm{NL}$-type trispectra in the equilateral momentum configuration where the signal is expected to be maximized in inflationary models with an axion-gauge coupling.

This paper is organized as follows.
In Sec.~\ref{sec: model and background}, we introduce the setup of the model and briefly summarize the background dynamics.
In Sec.~\ref{sec: perturbation}, we present the fourth-order Lagrangian for the tensor perturbations and derive the third-order perturbations.
Then we use these perturbations to obtain the $g_\mathrm{NL}$-type GW trispectra in Sec.~\ref{sec: trispectrum}.
In Sec.~\ref{sec: gNL evaluation}, we evaluate the magnitude of the GW trispectra and discuss the observability for the available model parameters.
Sec.~\ref{sec: discussion} is devoted to the summary and discussion of our results.

\section{Model and Background dynamics}
\label{sec: model and background}

We consider the Lagrangian
\begin{equation}
    \mathcal{L}
    =
    \frac{M_\mathrm{P}}{2}R 
    + \mathcal{L}_\phi
    + \frac{1}{2}\partial_\mu \chi \partial^\mu \chi 
    - V(\chi)
    -\frac{1}{4} F^a_{\mu \nu} F^{a \mu \nu}
    +\frac{\lambda}{4f}\chi F^a_{\mu \nu} \tilde{F}^{a\mu \nu},
\end{equation}
where $M_\mathrm{P}$ is the reduced Planck mass,
$\mathcal{L}_\phi$ is the Lagrangian of the inflaton $\phi$,
$V(\chi)$ is the potential of the axion field $\chi$, 
$\lambda$ is a dimensionless coupling constant of the Chern-Simons term,
and $f$ is a decay constant of the axion field.
In this paper, we do not specify $\mathcal{L}_\phi$ and $V(\chi)$.
The field strength of a SU(2) gauge field $F^a_{\mu \nu}$ is defined by
\begin{equation}
    F^a_{\mu \nu} 
    \equiv
    \partial_\mu A^a_\nu - \partial_\nu A^a_\mu - g \epsilon^{a b c} A^b_\mu A^c_\nu,
\end{equation}
and its dual $\tilde{F}^{a\mu\nu}$ is defined by
\begin{equation}
    \tilde{F}^{a\mu \nu} 
    \equiv 
    \frac{\epsilon^{\mu \nu \rho \sigma}}{2\sqrt{-\tilde{g}}} F^a_{\rho \sigma},
\end{equation}
where $g$ is the coupling constant of the SU(2) gauge group,
and $\tilde{g}$ is the determinant of the space-time metric $g_{\mu\nu}$.
For anti-symmetric tensors, we use
\begin{equation}
    \epsilon^{123} = 1, \quad
    \epsilon^{0123} = 1.
\end{equation}

We assume that the background axion field $\chi_0$ slowly rolls down $V(\chi)$.
Under this assumption, the background gauge field has an attractor configuration~\cite{Adshead:2012kp,Maleknejad:2013npa,Domcke:2018rvv}:
\begin{equation}
    \bar{A}^a_0 = 0, \quad
    \bar{A}^a_i = \delta^a_i a(t)Q(t),
\end{equation}
where $a(t)$ is the scale factor.
This solution respects the isotropy of the Universe~\cite{Maleknejad:2013npa,Wolfson:2020fqz,Wolfson:2021fya} and the background metric can be denoted as
\begin{equation}
    \mathrm{d}s^2 
    =
    g_{\mu\nu} \mathrm{d}x^{\mu}\mathrm{d}x^{\nu}
    =
    \mathrm{d}t^2 - a(t)^2(\mathrm{d}x^2 + \mathrm{d}y^2 + \mathrm{d}z^2).
\end{equation}

We decompose the axion and the gauge fields into the background and the perturbation as
\begin{equation}
    \chi(t, \bm x) = \chi_0(t) + \delta\chi(t, \bm x), \quad
    A^a_i(t, \bm x) = \delta^a_i a(t) Q(t) + \delta A^a_i(t, \bm x).
\end{equation}
We will discuss the non-dynamical component $\delta A^a_0$ later.
The equations of motion (EoMs) of the background fields are given by
\begin{align}
    \ddot{\chi}_0 + 3H\dot{\chi}_0 + \partial_\chi V(\chi_0)
    &= 
    -\frac{3g\lambda}{f}Q^2\left( \dot{Q} + HQ \right),
    \label{eq: chi BG EoM}
    \\
    \ddot{Q} + 3H\dot{Q} + \left( \dot{H} + 2H^2 \right) Q + 2g^2 Q^3
    &=
    \frac{g\lambda}{f}Q^2\dot{\chi}_0,
    \label{eq: Q BG EoM}
\end{align}
where the dots represent time derivatives $\partial_t$,
and $H \equiv \dot{a}/a$ is the Hubble parameter.
The right hand side of Eq.~\eqref{eq: chi BG EoM} slows down the time evolution of $\chi_0$ in addition to the Hubble friction term.
Here we introduce two dimensionless parameters:
\begin{equation}
    m_Q(t) 
    \equiv
    \frac{g Q}{H}, \quad
    \Lambda(t)
    \equiv
    \frac{\lambda Q}{f}.
\end{equation}
In the slow-roll regime, $m_Q \gtrsim 1$ and $\Lambda \gg 1$,
$Q$ is stabilized by its effective mass, and $\chi_0$ is significantly slowed down by the coupling to the gauge field.
By dropping the time derivative terms in Eq.~\eqref{eq: chi BG EoM}, we obtain
\begin{equation}
    m_Q 
    \simeq
    \left(
        \frac{-g^2 f \partial_\chi V(\chi_0)}{3\lambda H^4}
    \right)^{1/3},
\end{equation}
and, from Eq.~\eqref{eq: Q BG EoM}, we obtain the time derivative of the axion background field as
\begin{equation}
    \xi 
    \equiv
    \frac{\lambda \dot{\chi}_0}{2f H}
    \simeq
    m_Q + m_Q^{-1}.
\end{equation}

The Einstein equations of the background are given by
\begin{align}
    3M_\mathrm{P} H^2 
    &=
    \rho_\phi + \frac{1}{2}\dot{\chi}_0^2 + V(\chi_0) 
    + \frac{3}{2}\left( \dot{Q}^2 + HQ \right)^2 + \frac{3}{2}g^2Q^4,
    \\
    -\frac{\dot{H}}{H^2}
    &=
    \epsilon_\phi + \epsilon_\chi + \epsilon_B + \epsilon_E,
\end{align}
where $\rho_\phi$ is the energy density of the inflaton and the slow-roll parameters are defined by
\begin{align}
    \epsilon_\phi \equiv -\frac{\dot{\rho_\phi}}{6M_\mathrm{P}^2 H^3},
    \qquad
    \epsilon_\chi \equiv \frac{\dot{\chi}_0^2}{2M_\mathrm{P}^2 H^2},
    \qquad
    \epsilon_E \equiv \frac{\left( \dot{Q} + HQ\right)^2}{M_\mathrm{P}^2 H^2},
    \qquad
    \epsilon_B \equiv \frac{g^2Q^4}{M_\mathrm{P}^2H^2}.
\end{align}
In the dynamics of the perturbations, the background fields appear only through $H, m_Q, \xi, \epsilon_E$, and $\epsilon_B$.
By using the relations among these quantities, we can eliminate $\xi$ and $\epsilon_E$ in the slow-roll regime, $\dot{Q} \ll HQ$.
In addition, we can ignore the time variations of $H, m_Q$, and $\epsilon_B$ in the leading order approximation in the slow-roll regime.
In other words, we can characterize the background dynamics by three constants $H, m_Q$, and $\epsilon_B$ as in the following discussion of the perturbations.

\section{GW perturbations}
\label{sec: perturbation}

In this section, we investigate the tensor perturbation.
In order to calculate the trispectrum of GWs, we need to expand the action up to the fourth order in perturbations.
We define the tensor perturbations of the metric and the SU(2) gauge field as
\begin{equation}
    g_{i j} = -a^2(\delta_{i j} + h_{i j}), \quad
    \delta A^a_i = t_{a i} + \cdots,
\end{equation}
where $\cdots$ represents the scalar and vector perturbations, which we neglect in this paper.
We impose the transverse and traceless condition on $h_{i j}$ and $t_{i j}$.
Note that, although $t_{a i}$ is not a tensor in a strict sense, $t_{a i}$ transforms as a tensor in practice since the background gauge field $\bar{A}^a_i$ which is multiplied by $t_{a i}$ is proportional to $\delta^a_i$.
The inverse metric is given by
\begin{equation}
    g^{i j} 
    = 
    -a^{-2} \left( 
        \delta_{i j} - h_{i j} + h_{i k} h_{k j} + \mathcal{O}(h^3)
    \right).
\end{equation}
For later convenience, we define $\psi_{i j}$ as
\begin{equation}
    \psi_{i j} \equiv \frac{a M_\mathrm{P}}{2}h_{i j}.
\end{equation}

In the following, we use the conformal time $\tau$ instead of $t$:
\begin{equation}
    a \mathrm{d} \tau \equiv \mathrm{d} t,
\end{equation}
and the metric is
\begin{equation}
    \mathrm{d} s^2 
    =
    g_{\mu \nu} \mathrm{d}x^\mu \mathrm{d}x^\nu
    =
    a^2 \left[
        \mathrm{d} \tau^2 - (\delta_{i j} + h_{i j}) \mathrm{d}x^i \mathrm{d} x^j
    \right].
\end{equation}

We expand the action up to the fourth order in perturbations as 
\begin{equation}
    S_\mathrm{pert} 
    =
    \int \mathrm{d}\tau \mathrm{d}^3 x \,
    \left[ 
        L_2 + L_3 + L_4
    \right],
\end{equation}
where $L_i \, (i = 2,3,4)$ is the $i$-th order term in perturbations.
In the rest of this section, we discuss $L_2$ and $L_4$ in order, while we skip the discussion of $L_3$ for the following reason.
The trispectra has two types of the contributions: those generated through a single four-point vertex and those generated through two three-point vertices. The former and latter ones correspond to the $g_{\rm NL}$ type contribution, $\langle \psi^4 \rangle_{g_{\rm NL}} \sim \langle \psi_1^3 \psi_3 \rangle$, and the $\tau_{\rm NL}$ type one, $\langle \psi^4 \rangle_{\tau_{\rm NL}} \sim \langle \psi_1^2 \psi_2^2 \rangle$, respectively, where $\psi_n$ denotes the $n$-th order term of the GW. As mentioned at the end of Sec.~\ref{sec: introduction}, the $g_{\rm NL}$ type signal becomes a unique probe of the axion-SU(2) gauge coupling under examination. Furthermore, this is easier to test with the CMB data. Therefore, in the following, we focus on the contribution with a single four-point vertex
to obtain the $g_\mathrm{NL}$ type tensor trispectrum.

\subsection{Second order Lagrangian and First order perturbations}
\label{subsec: L2}

First, we expand the Lagrangian up to the second order in perturbations and derive the first order perturbations.
We divide $L_2$ into three contributions,
$L_2 = L_2^{\psi^2} + L_2^{\psi t} + L_2^{t^2}$:
\begin{align}
    L_2^{\psi^2}
    &=
    \frac{1}{2}\psi_{i j}' \psi_{i j}'
    -\frac{1}{2}\partial_k \psi_{i j} \partial_k \psi_{i j}
    + \frac{1}{\tau^2} \psi_{i j} \psi_{i j},
    \\
    L_2^{\psi t}
    &=
    \frac{2\sqrt{\epsilon_B}}{\tau}
    \left[
        \frac{1}{m_Q}\psi_{i j} t_{i j}' 
        + \psi_{i j} \epsilon^{ikl} \partial_l t_{jk}
        +\frac{m_Q}{\tau} \psi_{i j} t_{i j}
    \right],
    \\
    L_2^{t^2}
    &= 
    \frac{1}{2}t_{i j}'t_{i j}' -\frac{1}{2}\partial_k t_{i j} \partial_k t_{i j}
    +\frac{2m_Q + m_Q^{-1}}{\tau} \epsilon^{ijk} t_{i l} \partial_j t_{k l}
    -\frac{m_Q^2 + 1}{\tau^2} t_{i j} t_{i j},
\end{align}
where we used $\tau \simeq -1/(a H)$.
Then, the EoMs of $\psi_{i j}$ and $t_{i j}$ are given by
\begin{align}
    &\psi_{i j}'' - \partial_k \partial_k \psi_{i j} - \frac{2}{\tau^2} \psi_{i j}
    =
      \frac{2\sqrt{\epsilon_B}}{m_Q \tau} t_{i j}'
    + \frac{2\sqrt{\epsilon_B}}{\tau} \epsilon^{i k l} \partial_l t_{j k}
    + \frac{2\sqrt{\epsilon_B}m_Q}{\tau^2} t_{i j},
    \\
    &t_{i j}'' - \partial_k \partial_k t_{i j} 
    + \frac{2(2m_Q + m_Q^{-1})}{\tau} \epsilon^{i k l} \partial_l t_{j k}
    + \frac{2(m_Q^2 + 1)}{\tau^2} t_{i j}
    =
    \mathcal{O}(\psi_{i j}).
\end{align}
Since the tachyonic instability takes place only in $t_{i j}$, 
$\psi_{i j}$ is not amplified as much as $t_{i j}$ and then we ignore the linear term of $\psi_{i j}$ in the EoM for $t_{i j}$.

To see the amplification of $t_{i j}$,
we decompose $\psi_{i j}$ and $t_{i j}$ with the circular polarization tensors as
\begin{equation}
    X_{i j}(\tau, \bm{x})
    =
    \int \frac{\mathrm{d}^3 k}{(2\pi)^3} e^{i \bm{k} \cdot \bm{x}}
    \left[ 
          e_{i j}^R (\hat{\bm{k}}) X_{\bm{k}}^R (\tau)
        + e_{i j}^L (\hat{\bm{k}}) X_{\bm{k}}^L (\tau)
    \right],
\end{equation}
where $X = \psi, t$, and the definitions and properties of the polarization tensors are summarized in App.~\ref{app: polarization tensor}.
Note that the polarization tensors satisfy
$i \epsilon^{i k l} k^l e_{j k}^{R/L}(\hat{\bm{k}}) = \pm k e_{i j}^{R/L}(\hat{\bm{k}})$.

In addition, we quantize $\psi_{\bm{k}}^{R/L}(\tau)$ and $t_{\bm{k}}^{R/L}(\tau)$ and expand them in a perturbative series as
\begin{equation}
    \hat{X}_{\bm{k}}^p (\tau)
    =
    \hat{X}_1^p (\tau, \bm{k}) + \hat{X}_2^p (\tau, \bm{k}) + \cdots,
\end{equation}
where $p = R, L$ represents the polarization.
The first order components are written as
\begin{align}
    \hat{\psi}_1^p(\tau, \bm{k})
    &=
    \Psi_1^p (\tau, k) \hat{a}_{\bm{k}}^p
    + \Psi_1^{p*} (\tau, k) \hat{a}_{-\bm{k}}^{p \dagger},
    \\
    \hat{t}_1^p(\tau, \bm{k})
    &=
    T_1^p (\tau, k) \hat{a}_{\bm{k}}^p
    + T_1^{p*} (\tau, k) \hat{a}_{-\bm{k}}^{p \dagger},
\end{align}
where $\hat{a}_{\bm{k}}^p$ and $\hat{a}_{\bm{k}}^{p \dagger}$ are the creation and annihilation operators, which satisfy the canonical commutation relation:
\begin{equation}
    [ \hat{a}_{\bm{k}}^p, \hat{a}_{-\bm{k}'}^{q \dagger}]
    =
    (2\pi)^3 \delta^{p q} \delta^{(3)}(\bm{k} + \bm{k}').
\end{equation}
Since we consider only GWs sourced by the gauge fields in this paper, we assign the same quantum operator as $\hat{t}_1^p$ to $\hat{\psi}_1^p$ and ignore the intrinsic vacuum fluctuations of $\hat{\psi}^p$.
By using these expressions, we obtain the EoMs for the mode functions as
\begin{align}
    & \partial_\tau^2 \Psi_1^{R/L}
    + \left[
        k^2 - \frac{2}{\tau^2}
    \right] \Psi_1^{R/L}
    = 
    \frac{2 \sqrt{\epsilon_B}}{m_Q \tau} \partial_\tau T_1^{R/L}
    \pm \frac{2 k \sqrt{\epsilon_B}}{\tau} T_1^{R/L}
    + \frac{2 \sqrt{\epsilon_B} m_Q}{\tau^2} T_1^{R/L}, 
    \label{eq: Psi mode function EoM}
    \\
    & \partial_\tau^2 T_1^{R/L}
    + \left[
        k^2 \pm \frac{2 k (2 m_Q + m_Q^{-1})}{\tau} + \frac{2(m_Q^2 + 1)}{\tau^2}
    \right] T_1^{R/L}
    = 
    \mathcal{O} \left( \Psi_1^{R/L} \right).
    \label{eq: T mode function EoM}
\end{align}
Due to the sign of the term proportional to $\tau^{-1}$ in the LHS of Eq.~\eqref{eq: T mode function EoM}, only $T_1^R$ experiences a tachyonic instability and it sources only $\Psi_1^R$ through Eq.~\eqref{eq: Psi mode function EoM}.
Therefore, we only consider the right-handed polarization modes in the following.
We can analytically express the solution of Eq.~\eqref{eq: T mode function EoM} as
\begin{equation}
    T_1^R (\tau, k)
    =
    \frac{1}{\sqrt{2k}} e^{\pi(2m_Q + m_Q^{-1})/2} W_{\beta, \alpha} (2i k \tau),
    \label{eq: T1R Whittaker solution}
\end{equation}
where $W_{\beta,\alpha} (z)$ is the Whittaker function, $\alpha \equiv -i\sqrt{2m_Q^2 + 7/4}$, and $\beta \equiv -i(2m_Q + m_Q^{-1})$.
This solution satisfies the Bunch-Davies initial condition in the sub-horizon limit.
In other words, $T_1^R$ approaches $e^{-ik\tau}/\sqrt{2k}$ as $k\tau \to -\infty$.
We can also obtain $\Psi_1^R$ by using the Green's function method as
\begin{equation}
    \Psi_1^R(\tau, k)
    =
    \int_{-\infty}^\infty \mathrm{d}\eta\,
    G_\psi(\tau, \eta, k) \mathcal{D}(\eta, k) T_1^R(\eta, k),
    \label{eq: Psi1 Green integration}
\end{equation}
with
\begin{align}
    G_\psi(\tau, \eta, k)
    &\equiv
    \frac{\Theta(\tau - \eta)}{k^3 \tau \eta}
    \left[ 
        k(\eta - \tau) \cos\left( k(\tau-\eta) \right)
        + (1 + k^2\tau \eta) \sin\left( k(\tau - \eta) \right)
    \right],
    \\
    \mathcal{D}(\eta, k)
    &\equiv
    \frac{2\sqrt{\epsilon_B}}{m_Q \eta}\partial_\eta
    +\frac{2\sqrt{\epsilon_B}}{\eta^2}(m_Q + k\eta),
\end{align}
where $\Theta(x)$ is the unit Heaviside function.

In the super-horizon limit, we can obtain the analytical expression of $\Psi_1^R(\tau, k)$ from Eq.~\eqref{eq: Psi1 Green integration} as
\begin{equation}
    \lim_{|k\tau| \to 0} \Psi_1^R(\tau, k)
    =
    \frac{\sqrt{\epsilon_B}}{\sqrt{2k}k\tau} \mathcal{F}(m_Q),
    \label{eq: Psi1 super-horizon limit}
\end{equation}
where $|\mathcal{F}(m_Q)|\sim e^{2.4m_Q}$ and its exact expression is given in~\cite{Dimastrogiovanni:2016fuu}.
From this solution, we obtain the power spectrum of $h$ in the super-horizon limit as
\begin{equation}
    \frac{k^3}{2\pi^2} P_h^{\mathrm{sourced}}
    =
    \frac{\epsilon_B H^2}{\pi^2 M_P^2}| \mathcal{F}(m_Q)|^2.
\end{equation}
This result should be contrasted with the tensor power spectrum of the vacuum fluctuation in the super-horizon limit,
\begin{equation}
    \frac{k^3}{2\pi^2} P_h^{\mathrm{vac}}
    =
    \frac{2 H^2}{\pi^2 M_P^2}.
\end{equation}
Compared to $P_h^{\mathrm{vac}}$, $P_h^{\mathrm{sourced}}$ is suppressed by $\epsilon_B\ll1$
but exponentially enhanced by $|\mathcal{F}(m_Q)|^2$.
For later convenience, we define their tensor-to-scalar ratios,
\begin{equation}
    r_{\mathrm{vac}}= \frac{P_h^{\mathrm{vac}}}{P_\zeta},
    \qquad
    r_{\mathrm{src}}= \frac{P_h^{\mathrm{sourced}}}{P_\zeta},
\end{equation}
where $P_\zeta$ is the power spectrum of the curvature perturbation observed on the CMB scale.
The additional production of PGWs by the gauge fields significantly affects the tensor power spectrum in observation, unless $r_{\mathrm{src}}$ is negligibly smaller than $r_{\mathrm{vac}}$.

\subsection{Fourth order Lagrangian and Third order perturbations}
\label{subsec: L4}

Next, since we are interested in the trispectrum generated through a single four-point vertex, we 
skip the third order terms and concentrate on the fourth order terms.
The fourth-order term $L_4$ can be divided into five components according to the power of $\psi$.
By comparing the power of the Planck mass, we can pick up the components making dominant contributions to the GW trispectrum.

First, we compare two components in $L_4$, $L_4^{\psi t^3}$ and $L_4^{t^4}$, which contain the $\mathcal{O}(\psi t^3)$ and $\mathcal{O}(t^4)$ terms, respectively.
Since each $\psi$ introduces the suppression factor $M_\mathrm{P}^{-1}$, $L_4^{\psi t^3}$ is suppressed by one more $M_\mathrm{P}^{-1}$ than $L_4^{t^4}$.
On the other hand, $L_4^{\psi t^3}$ induces $\psi_3$ from three $t_1$'s, while $L_4^{t^4}$ induces $t_3$ from three $t_1$'s.
When we evaluate the GW trispectrum, $t_3$ have to be converted to $\psi_3$ through the linear relation between $\psi$ and $t$~\eqref{eq: Psi1 Green integration}, which introduces another suppression factor $\sqrt{\epsilon_B} \propto M_\mathrm{P}^{-1}$ to the contribution of $L_4^{t^4}$.
Therefore, the contributions to the GW trispectrum of $L_4^{\psi t^3}$ and $L_4^{t^4}$ have the same power of the Planck mass.

Next, we show that the contribution of $L_4^{\psi^2 t^2}$ is subdominant to that of $L_4^{\psi t^3}$.
$L_4^{\psi^2 t^2}$ induces $\psi_3$ from two $t_1$'s and $\psi_1$, while $L_4^{\psi t^3}$ induces $\psi_3$ from three $t_1$'s.
Since $\psi_1$ is suppressed by  $\sqrt{\epsilon_B} \propto M_\mathrm{P}^{-1}$ compared to $t_1$ through Eq.~\eqref{eq: Psi1 Green integration}, the difference in the sourcing perturbations brings this suppression factor to the contribution of $L_4^{\psi^2 t^2}$.
Moreover, $L_4^{\psi^2 t^2}$ itself is suppressed by one more $M_\mathrm{P}^{-1}$ than $L_4^{\psi t^3}$.
Therefore, the contribution of $L_4^{\psi^2 t^2}$ is suppressed by $M_\mathrm{P}^{-2}$ compared to that of $L_4^{\psi t^3}$.
In the similar way, we can see that the $\mathcal{O}(\psi^4)$ and $\mathcal{O}(\psi^3 t)$ terms also make subdominant contributions.
Thus, we discuss only $L_4^{\psi t^3}$ and $L_4^{t^4}$ in the following.

In the calculation of the fourth order perturbation, we have to take into account the non-dynamical variables of the gauge field $\delta A_0^a$, 
which is decomposed into a scalar and an effective vector component as
\begin{align}
    \delta A_0^a = a^{-1}(t) [\partial_a Y + Y_a].
\end{align}
Note that the subscript $0$ corresponds to not $\tau$ but $t$.
Since $\delta A_0$ does not include tensor components,
it does not couple to linear tensor perturbations but couples to quadratic tensor perturbations
(e.g. $\delta A_0 t^2$) in the expanded Lagrangian.
Therefore, the integration out of $\delta A_0$ induces only the fourth or higher order interactions of tensor perturbations.

\subsubsection{Contribution without non-dynamical variables}
\label{subsubsec: dynamical}

First, we consider $L_4$ without the contribution from the non-dynamical variables.
After some calculations, we obtain
\begin{align}
    L_4^{\psi t^3}
    &=
    c^{\psi t^3} \psi_{i j} \left[
        t_{i j} t_{k l} t_{k l} - 2t_{j k} t_{k l} t_{l i}
        +\frac{\tau}{m_Q} \epsilon^{a b c} 
        (\partial_l t_{a i}- \partial_i t_{a l}) t_{b l} t_{c j}
    \right],
    \\
    L_4^{t^4}
    &=
    \frac{c^{t^4}}{4}(t_{i j} t_{j k} t_{k l} t_{l i} - t_{i j} t_{i j} t_{k l} t_{k l}),
\end{align}
where
\begin{equation}
    c^{\psi t^3} 
    \equiv
    \frac{2m_Q^3 H^2}{\sqrt{\epsilon_B}M_\mathrm{P}^2}
    =
    g c^{\psi t^2},
    \quad
    c^{t^4} 
    \equiv
    g^2
    =
    \frac{m_Q^4 H^2}{\epsilon_B M_\mathrm{P}^2} = g c^{t^3},
\end{equation}
with the coefficients of the third order vertices~\cite{Agrawal:2018mrg}
\begin{equation}
    c^{\psi t^2} \equiv \frac{2 m_Q H}{M_\mathrm{P}},
    \quad
    c^{t^3} \equiv g = \frac{m_Q^2 H}{\sqrt{\epsilon_B} M_\mathrm{P}}. 
\end{equation}
Note that these coefficients satisfy a hierarchical relation,
\begin{equation}
    \frac{c^{\psi t^3}}{c^{t^4}}
    =
    \frac{c^{\psi t^2}}{c^{t^3}}
    =
    \frac{2\sqrt{\epsilon_B}}{m_Q}
    =
    \frac{2Q}{M_\mathrm{P}}
    \ll
    1.
\end{equation}

After the Fourier transformation, we obtain
\begin{align}
    S_4^{\psi t^3}
    =&
    c^{\psi t^3} \int \frac{\mathrm{d} \tau \mathrm{d}^3 k \mathrm{d}^3 p \mathrm{d}^3 q \mathrm{d}^3 r}{(2\pi)^9}
    \delta^{(3)}(\bm{k} + \bm{p} + \bm{q} + \bm{r})
    \psi_{\bm{k}}^R t_{\bm{p}}^R t_{\bm{q}}^R t_{\bm{r}}^R
    \nonumber\\
    & \times e_{i j}^R(\hat{\bm{k}})
    \left[
        e_{i j}^R(\hat{\bm{p}}) e_{k l}^R(\hat{\bm{q}}) e_{k l}^R(\hat{\bm{r}})
        -2 e_{j k}^R(\hat{\bm{p}}) e_{k l}^R(\hat{\bm{q}}) e_{l i}^R(\hat{\bm{r}})
        + \frac{\tau}{m_Q} \epsilon^{a b c} i 
        \left(
            p^l e_{a i}^R(\hat{\bm{p}}) - p^i e_{a l}^R(\hat{\bm{p}})
        \right)
        e_{b l}^R(\hat{\bm{q}}) e_{c j}^R(\hat{\bm{r}})
    \right],
    \label{eq: S4 psi t3 first expression}
    \\
    S_4^{t^4}
    =&
    c^{t^4} \int \frac{\mathrm{d} \tau \mathrm{d}^3 k \mathrm{d}^3 p \mathrm{d}^3 q \mathrm{d}^3 r}{(2\pi)^9}
    \delta^{(3)}(\bm{k} + \bm{p} + \bm{q} + \bm{r})
    t_{\bm{k}}^R t_{\bm{p}}^R t_{\bm{q}}^R t_{\bm{r}}^R
    \nonumber\\
    & \times e_{i j}^R(\hat{\bm{k}})
    \left(
        e_{j k}^R(\hat{\bm{p}}) e_{k l}^R(\hat{\bm{q}}) e_{l i}^R(\hat{\bm{r}})
        - e_{i j}^R(\hat{\bm{p}}) e_{k l}^R(\hat{\bm{q}}) e_{k l}^R(\hat{\bm{r}})
    \right),
    \label{eq: S4 t4 first expression}
\end{align}
where we represent $\psi^R(\tau, \bm{k})$ and $t^R(\tau, \bm{k})$ by $\psi^R_{\bm{k}}$ and $t^R_{\bm{k}}$, respectively.

\subsubsection{Contribution from non-dynamical variables}
\label{subsubsec: non-dynamical}

Next we consider the contribution from the non-dynamical variable $Y$ and $Y_a$.
By completing the square and integrating out $Y$ and $Y_a$,
we obtain the fourth order Lagrangian coming from the non-dynamical components.

As shown in App.~\ref{App: 4th Lagrangian from non-dynamical}, the $\mathcal{O}(\psi t^3)$ and $\mathcal{O}(t^4)$ terms coming from the non-dynamical scalar components are
\begin{align}
    S_{\mathrm{nd, s}}^{\psi t^3}
    =
    -\int 
    &
    \frac{\mathrm{d}\tau \mathrm{d}^3 k \mathrm{d}^3 p \mathrm{d}^3 q \mathrm{d}^3 r}{2(2\pi)^9}
    \delta^{(3)}( \bm{k} + \bm{p} + \bm{q} + \bm{r})
    F_\mathrm{s}(\tau, |\bm{k} + \bm{p}|)
    \nonumber\\
    &   
    \times \psi_{\bm{k}}
    \left(
        \frac{\tau}{m_Q} t_{\bm{p}}'
        S_\mathrm{s}(\bm{p},\bm{k})
        -
        \frac{c^{\psi t^2}}{c^{t^3}}
        \left(t_{\bm{p}}' + \frac{t_{\bm{p}}}{\tau}\right)
        A_\mathrm{s}(\bm{p},\bm{k})
    \right)
    t_{\bm{q}}' t_{\bm{r}}
    A_\mathrm{s}(\bm{q}, \bm{r}),
    \label{eq: Ss psi t3}
    \\
    S_{\mathrm{nd, s}}^{t^4}
    =
    -\int 
    &
    \frac{\mathrm{d}\tau \mathrm{d}^3 k \mathrm{d}^3 p \mathrm{d}^3 q \mathrm{d}^3 r}{4(2\pi)^9}
    \delta^{(3)}( \bm{k} + \bm{p} + \bm{q} + \bm{r})
    F_\mathrm{s}(\tau, |\bm{k}+ \bm{p}|)
    t_{\bm{p}}' t_{\bm{k}}
    t_{\bm{q}}' t_{\bm{r}}
    A_\mathrm{s}(\bm{p}, \bm{k})
    A_\mathrm{s}(\bm{q}, \bm{r}),   
    \label{eq: Ss t4}
\end{align}
where
\begin{align}
    F_\mathrm{s}(\tau, |\bm{k}+ \bm{p}|) 
    & \equiv
    \left( 
        \frac{|\bm{k} + \bm{p}|^4}{2} 
        + \frac{m_Q^2 |\bm{k} + \bm{p}|^2}{\tau^2} 
    \right)^{-1},
    \\
    S_\mathrm{s}(\bm{p}, \bm{k})
    & \equiv
    c^{\psi t^2} (k^j + p^j) (k^k + p^k) 
    e_{i j}^R(\hat{\bm{p}}) e_{i k}^R({\hat{\bm{k}}}).
    \\
    A_\mathrm{s}(\bm{p}, \bm{k})
    & \equiv
    c^{t^3}(p - k) 
    e_{i j}^R(\hat{\bm{p}}) e_{i j}^R(\hat{\bm{k}}).
\end{align}

Moreover, the non-dynamical vector components induce the $\mathcal{O}(\psi t^3)$ and $\mathcal{O}(t^4)$ terms as
\begin{align}
    S_{\mathrm{nd, v}}^{\psi t^3}
    =
    - \int 
    &
    \frac{\mathrm{d} \tau \mathrm{d}^3 k \mathrm{d}^3 p \mathrm{d}^3 q \mathrm{d}^3 r}
    {2 (2\pi)^9}
    \delta^{(3)}(\bm{k} + \bm{p} + \bm{q} + \bm{r})
    \sum_{\lambda = R, L}
    F_\mathrm{v}^\lambda(\tau, |\bm{k}+\bm{p}|)
    \nonumber\\
    & \times
    \psi_{\bm{k}}
    \left(
        \frac{\tau}{m_Q} P_\mathrm{v}^\lambda(\bm{p},\bm{k}) t'_{\bm{p}}
        -
        \frac{c^{\psi t^2}}{c^{t^3}}
        A_\mathrm{v}^\lambda(\bm{p},\bm{k})
        \left[ t'_{\bm{p}} + \frac{t_{\bm{p}}}{\tau}\right]
    \right)
    A_\mathrm{v}^\lambda(\bm{q},\bm{r})
    t_{\bm{q}} t'_{\bm{r}},
    \label{eq: Sv psi t3}
    \\
    S_{\mathrm{nd, v}}^{t^4}
    =
    -\int 
    &
    \frac{\mathrm{d}\tau \mathrm{d}^3 k \mathrm{d}^3 p \mathrm{d}^3 q \mathrm{d}^3 r}{4(2\pi)^9}
    \delta^{(3)}( \bm{k} + \bm{p} + \bm{q} + \bm{r})
    \sum_{\lambda = R, L}
    F^\lambda_\mathrm{v}(\tau, |\bm{k}+ \bm{p}|)
    A^\lambda_\mathrm{v}(\bm{p}, \bm{k}) t_{\bm{p}} t_{\bm{k}}'
    A^\lambda_\mathrm{v}(\bm{q}, \bm{r}) t_{\bm{q}} t_{\bm{r}}',    
    \label{eq: Sv t4}
\end{align}
where
\begin{align}
    F^\lambda_\mathrm{v}(\tau, |\bm{k}+ \bm{p}|) 
    & \equiv 
    \left(
        \frac{|\bm{k}+\bm{p}|^2}{2} + s_\lambda |\bm{k}+\bm{p}| \frac{m_Q}{\tau} + \frac{m_Q^2}{\tau^2}
    \right)^{-1},
    \\
    P^\lambda_\mathrm{v}(\bm{p}, \bm{k})
    & \equiv
    e_a^\lambda(\widehat{-\bm{k}-\bm{p}})
    c^{\psi t^2} i p^b
    e_{b c}^R(\hat{\bm{k}}) e_{a c}^R(\hat{\bm{p}}),
    \\
    A^\lambda_\mathrm{v}(\bm{p}, \bm{k})
    & \equiv
    e_a^\lambda(\widehat{-\bm{k}-\bm{p}})
    c^{t^3} \epsilon^{a b c} 
    e_{b d}^R(\hat{\bm{k}}) e_{c d}^R(\hat{\bm{p}}),
\end{align}
with $s_{R/L} = \pm 1$.

\subsubsection{Third order perturbations}
\label{subsubsec: psi3 and t3}

Based on those expanded Lagrangian, we compute
the third order perturbation $\hat{t}_3$ and $\hat{\psi}_3$ coming from the four-point vertices.
The EoM for $\hat{t}_3$ reads
\begin{equation}
    \hat{t}_3^{R \prime \prime}(\tau, \bm{k})
    + \left(
        k^2 + \frac{2k(2m_Q + m_Q^{-1})}{\tau} + \frac{2(m_Q^2 + 1)}{\tau^2}
    \right) \hat{t}_3^R (\tau, \bm{k})
    =
    (2\pi)^3
    \left. 
        \overline{\frac{\delta S^{t^4}}{\delta \hat{t}^R(\tau, -\bm{k})}}
    \right|_{\hat{t}^R = \hat{t}_1^R},
\end{equation}
where $S^{t^4} \equiv S_4^{t^4} + S_{\mathrm{nd, s}}^{t^4} + S_{\mathrm{nd, v}}^{t^4}$, and $\hat{t}^R$ in the RHS is evaluated as $\hat{t}_1^R$.
Note that the source term $\overline{\delta S^{t^4}/\delta \hat{t}^R(\eta, -\bm{k})}$ is symmetrized with respect to the permutation of $\hat{t}^R$ in the quantization,
which is, in this case, equivalent to the quantization using the normal ordering.
This procedure is necessary to make the correlation functions independent from the order of the operators.

The solution of this EoM can be obtained by using the Green's function as 
\begin{equation}
    \hat{t}_{3,t_1^3}^R(\tau, \bm{k})
    =
    (2\pi)^3
    \int^\infty_{-\infty} \mathrm{d} \eta \, G_t (\tau, \eta, k)
    \left. 
        \overline{\frac{\delta S^{t^4}}{\delta \hat{t}^R(\eta, -\bm{k})}}
    \right|_{\hat{t}^R = \hat{t}_1^R}.
    \label{eq: t3 from three t1}
\end{equation}

As for each of the three terms in $S^{t^4}$, the source term is
\begin{align}
    (2 \pi)^3 \frac{\delta S_4^{t^4}}{\delta \hat{t}^R(\tau, -\bm{k})}
    =
    &
    c^{t^4} \int \frac{\mathrm{d}^3 p \mathrm{d}^3 q \mathrm{d}^3 r}{(2\pi)^6}
    \delta^{(3)}(-\bm{k} + \bm{p} + \bm{q} + \bm{r})
    \nonumber\\
    & \times 
    \left[
        e_{i j}^R(-\hat{\bm{k}}) e_{j k}^R(\hat{\bm{p}}) e_{k l}^R(\hat{\bm{q}}) e_{l i}^R(\hat{\bm{r}})
        - e_{i j}^R(-\hat{\bm{k}}) e_{i j}^R(\hat{\bm{p}}) e_{k l}^R(\hat{\bm{q}}) e_{k l}^R(\hat{\bm{r}})
    \right]
    \hat{t}^R_{\bm{p}} \hat{t}^R_{\bm{q}} \hat{t}^R_{\bm{r}},
    \label{eq: t3 source dynamical}
    \\
    (2 \pi)^3 \frac{\delta S_{\mathrm{nd, s}}^{t^4}}{\delta \hat{t}^R(\tau, -\bm{k})}
    =&
    -\int \frac{\mathrm{d}^3 p \mathrm{d}^3 q \mathrm{d}^3 r}{2(2\pi)^6}
    \delta^{(3)}(-\bm{k} + \bm{p} + \bm{q} + \bm{r})
    A_\mathrm{s}(\bm{p}, -\bm{k}) A_\mathrm{s}(\bm{q}, \bm{r})
    \nonumber\\
    &
    \times 
    \left[
        F_\mathrm{s}(\tau, |-\bm{k}+ \bm{p}|)
        \hat{t}^{R \prime}_{\bm{p}}
        \hat{t}^{R \prime}_{\bm{q}}
        \hat{t}^R_{\bm{r}}
        + \left\{
            F_\mathrm{s}(\tau, |-\bm{k}+ \bm{p}|)
            \hat{t}^R_{\bm{p}}
            \hat{t}^{R \prime}_{\bm{q}}
            \hat{t}^R_{\bm{r}}
        \right\}'
    \right],
    \label{eq: t3 source scalar}
    \\
    (2 \pi)^3 \frac{\delta S_{\mathrm{nd, v}}^{t^4}}{\delta \hat{t}^R(\tau, -\bm{k})}
    =&
    -\int \frac{\mathrm{d}^3 p \mathrm{d}^3 q \mathrm{d}^3 r}{2(2\pi)^6}
    \delta^{(3)}(-\bm{k} + \bm{p} + \bm{q} + \bm{r})
    \sum_{\lambda = R,L}
    \nonumber\\
    &
    \times 
    \left[
        F^\lambda_\mathrm{v}(\tau, |-\bm{k}+ \bm{p}|)
        A^\lambda_\mathrm{v}(-\bm{k}, \bm{p})
        \hat{t}^{R \prime}_{\bm{p}}
        A^\lambda_\mathrm{v}(\bm{q}, \bm{r})
        \hat{t}^R_{\bm{q}} \hat{t}^{R \prime}_{\bm{r}}
    \right.
    \nonumber\\
    & \quad
    \left.
        - A^\lambda_\mathrm{v}(\bm{p}, -\bm{k})
        \left\{
            F^\lambda_\mathrm{v}(\tau, |-\bm{k}+ \bm{p}|)
            \hat{t}^R_{\bm{p}}
            A^\lambda_\mathrm{v}(\bm{q}, \bm{r})
            \hat{t}^R_{\bm{q}} \hat{t}^{R \prime}_{\bm{r}}
        \right\}'
    \right] \ .
    \label{eq: t3 source vector}
\end{align}
Next $\hat{\psi}_3$ is contributed by the following two processes:
$\hat{t}_3$ contributes through the linear coupling between $\psi$ and $t$,
and $\hat{t}_1^3$ does through the $\psi t^3$ vertex.
We can obtain the former contribution in a similar way as $\Psi_1^R$ in Eq.~\eqref{eq: Psi1 Green integration}:
\begin{equation}
    \hat{\psi}_{3,t_{3,t_1^3}}^R(\tau, \bm{k})
    =
    \int_{-\infty}^\infty \mathrm{d}\eta\,
    G_\psi(\tau, \eta, k) \mathcal{D}(\eta, k) \hat{t}_{3,t_1^3}^R(\eta, \bm{k}).
    \label{eq: psi3 from t3}
\end{equation}

As for the latter contribution, we consider the relevant EoM:
\begin{equation}
    \left(
        \partial_\tau^2 + k^2 - \frac{2}{\tau^2}
    \right) \hat{\psi}^R(\tau, \bm{k})
    =
    (2\pi)^3
    \left. 
        \overline{\frac{\delta S^{\psi t^3}}{\delta \hat{\psi}^R(\tau, -\bm{k})}}
    \right|_{\hat{t}^R = \hat{t}_1^R},
\end{equation}
where $S^{\psi t^3} \equiv S_4^{\psi t^3} + S_{\mathrm{nd, s}}^{\psi t^3} + S_{\mathrm{nd, v}}^{\psi t^3}$,
$\hat{t}^R$ in the RHS is evaluated as $\hat{t}_1^R$,
and $\overline{\delta S^{\psi t^3}/\delta \hat{\psi}^R(\eta, -\bm{k})}$ is symmetrized with respect to the permutation of $\hat{t}^R$.

The solution of this EoM can be obtained by using the Green's function as 
\begin{equation}
    \hat{\psi}_{3,t_1^3}^R(\tau, \bm{k})
    =
    (2\pi)^3
    \int^\infty_{-\infty} \mathrm{d} \eta \, G_\psi (\tau, \eta, k)
    \left. 
        \overline{\frac{\delta S^{\psi t^3}}{\delta \hat{\psi}^R(\eta, -\bm{k})}}
    \right|_{\hat{t}^R = \hat{t}_1^R}.
    \label{eq: psi3 from three t1}
\end{equation}

As for each of the three terms in $S^{\psi t^3}$, the source term is
\begin{align}
    (2 \pi)^3 \frac{\delta S_4^{\psi t^3}}{\delta \hat{\psi}^R(\tau, -\bm{k})}
    =&
    c^{\psi t^3} \int \frac{\mathrm{d}^3 p \mathrm{d}^3 q \mathrm{d}^3 r}{(2\pi)^6}
    \delta^{(3)}(-\bm{k} + \bm{p} + \bm{q} + \bm{r})
    t_{\bm{p}}^R t_{\bm{q}}^R t_{\bm{r}}^R
    \nonumber\\
    & \times e_{i j}^R(-\hat{\bm{k}})
    \left[
        e_{i j}^R(\hat{\bm{p}}) e_{k l}^R(\hat{\bm{q}}) e_{k l}^R(\hat{\bm{r}})
        -2 e_{j k}^R(\hat{\bm{p}}) e_{k l}^R(\hat{\bm{q}}) e_{l i}^R(\hat{\bm{r}})
        + \frac{\tau}{m_Q} \epsilon^{a b c} i 
        \left(
            p^l e_{a i}^R(\hat{\bm{p}}) - p^i e_{a l}^R(\hat{\bm{p}})
        \right)
        e_{b l}^R(\hat{\bm{q}}) e_{c j}^R(\hat{\bm{r}})
    \right],
    \label{eq: psi3 source dynamical}
    \\
    (2 \pi)^3 
    \frac{\delta S_\mathrm{nd, s}^{\psi t^3}}
    {\delta \hat{\psi}^R(\tau, -\bm{k})}
    =
    -\int 
    &
    \frac{\mathrm{d}^3 p \mathrm{d}^3 q \mathrm{d}^3 r}{2(2\pi)^6}
    \delta^{(3)}( -\bm{k} + \bm{p} + \bm{q} + \bm{r})
    F_\mathrm{s}(\tau, |\bm{p} - \bm{k}|)
    A_\mathrm{s}(\bm{q}, \bm{r})
    \nonumber\\
    &   
    \times
    \left(
        \frac{\tau}{m_Q} t_{\bm{p}}'
        S_\mathrm{s}(\bm{p},-\bm{k})
        -
        \frac{c^{\psi t^2}}{c^{t^3}}
        \left(t_{\bm{p}}' + \frac{t_{\bm{p}}}{\tau}\right)
        A_\mathrm{s}(\bm{p},-\bm{k})
    \right)
    t_{\bm{q}}' t_{\bm{r}},
    \label{eq: psi3 source scalar}
    \\
    (2 \pi)^3 
    \frac{\delta S_\mathrm{nd, v}^{\psi t^3}}
    {\delta \hat{\psi}^R(\tau, -\bm{k})}
    =
    - \int 
    &
    \frac{\mathrm{d}^3 p \mathrm{d}^3 q \mathrm{d}^3 r}
    {2 (2\pi)^6}
    \delta^{(3)}(-\bm{k} + \bm{p} + \bm{q} + \bm{r})
    \sum_{\lambda = R, L}
    F_\mathrm{v}^\lambda(\tau, |\bm{p}-\bm{k}|)
    A_\mathrm{v}^\lambda(\bm{q},\bm{r})
    \nonumber\\
    & \times
    \left(
        \frac{\tau}{m_Q} P_\mathrm{v}^\lambda(\bm{p},-\bm{k}) t'_{\bm{p}}
        -
        \frac{c^{\psi t^2}}{c^{t^3}}
        A_\mathrm{v}^\lambda(\bm{p},-\bm{k})
        \left[ t'_{\bm{p}} + \frac{t_{\bm{p}}}{\tau}\right]
    \right)
    t_{\bm{q}} t'_{\bm{r}}.
    \label{eq: psi3 source vector}
\end{align}

Now, we have the perturbations required to evaluate the $g_\mathrm{NL}$-type GW trispectra.

\begingroup
\allowdisplaybreaks[1]
\section{GW trispectrum}
\label{sec: trispectrum}

In this section, we calculate the tensor trispectrum of the right-handed GWs, $T_h^{RRRR}$, in the super-horizon limit:
\begin{align}
    (2\pi)^3 \delta^{(3)}( \bm{k}_1 + \bm{k}_2 + \bm{k}_3 + \bm{k}_4)
    T_h^{RRRR}(\bm{k}_1, \bm{k}_2, \bm{k}_3, \bm{k}_4)
    &\equiv
    \lim_{\tau \to 0} 
    \left\langle
        \hat{h}^R(\tau, \bm{k}_1 ) \hat{h}^R(\tau, \bm{k}_2 )
        \hat{h}^R(\tau, \bm{k}_3 ) \hat{h}^R(\tau, \bm{k}_4 )
    \right\rangle
    \nonumber\\
    &=
    \left( \frac{2}{a M_\mathrm{P}} \right)^4
    \lim_{\tau \to 0} 
    \left\langle
        \hat{\psi}^R(\tau, \bm{k}_1 ) \hat{\psi}^R(\tau, \bm{k}_2 )
        \hat{\psi}^R(\tau, \bm{k}_3 ) \hat{\psi}^R(\tau, \bm{k}_4 )
    \right\rangle.
\end{align}
The connected four-point correlator of the right-handed GW $\hat{\psi}^R = \hat{\psi}^R_1 + \hat{\psi}^R_2 + \hat{\psi}^R_3$ can be written as
\begin{align}
    \left\langle
        \hat{\psi}^R(\tau, \bm{k}_1 ) \hat{\psi}^R(\tau, \bm{k}_2 )
        \hat{\psi}^R(\tau, \bm{k}_3 ) \hat{\psi}^R(\tau, \bm{k}_4 )
    \right\rangle
    =&
    \left\langle
        \hat{\psi}_1^R(\tau, \bm{k}_1 ) \hat{\psi}_1^R(\tau, \bm{k}_2 )
        \hat{\psi}_2^R(\tau, \bm{k}_3 ) \hat{\psi}_2^R(\tau, \bm{k}_4 )
    \right\rangle
    +
    (\mathrm{permutation \, of \,} \hat{\psi}^R_1 \mathrm{\, and \,} \hat{\psi}^R_2)
    \nonumber\\
    +&
    \left\langle
        \hat{\psi}_1^R(\tau, \bm{k}_1 ) \hat{\psi}_1^R(\tau, \bm{k}_2 )
        \hat{\psi}_1^R(\tau, \bm{k}_3 ) \hat{\psi}_3^R(\tau, \bm{k}_4 )
    \right\rangle
    +
    (\mathrm{permutation \, of \,} \hat{\psi}^R_1 \mathrm{\, and \,} \hat{\psi}^R_3),
\end{align}
at leading order.
We are interested in the contributions from the four-point vertices,
which correspond to the second line in the above equation.
Such contributions can be divided into two terms according to the two types of $\hat{\psi}^R$, namely $\hat{\psi}^R_{3,t_{3,t_1^3}}$ and $\hat{\psi}^R_{3,t_1^3}$ as shown in Fig.~\ref{fig: diagrams}.
We will consider each term in the following.

\begin{figure}[htpb]
    \subfigure{\includegraphics[clip, width=0.25\columnwidth]{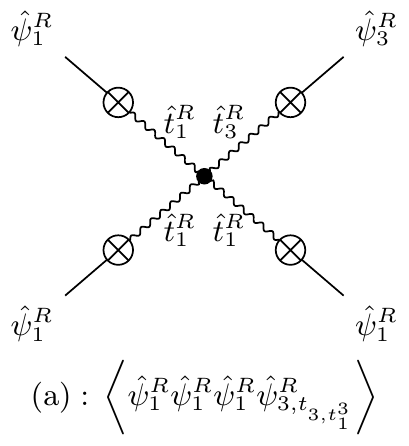}}
    \hspace{2.5cm}
    \subfigure{\includegraphics[clip, width=0.25\columnwidth]{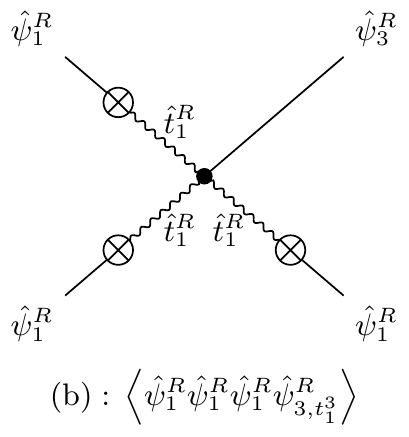}}
\caption{
Feynman diagrams illustrating the tree-level contributions to the $g_\mathrm{NL}$-type trispectrum of GWs.
The straight and wavy lines denote $\hat{\psi}^R$ and $\hat{t}^R$, respectively.
The black dots denote the vertices of the four-point interactions,
while the circled crosses denote the mixing between $\psi_{i j}$ and $t_{i j}$ through Eq.~\eqref{eq: Psi mode function EoM}.
}
\label{fig: diagrams}
\end{figure}

\subsection{diagram (a)}
\label{subsec: diagram a}

First, we consider diagram (a): 
$\left \langle \hat{\psi}_1^R(\tau, \bm{k}_1) \hat{\psi}_1^R(\tau, \bm{k}_2)  \hat{\psi}_1^R(\tau, \bm{k}_3) \hat{\psi}_{3,t_{3,t_1^3}}^R(\tau, \bm{k}_4)\right \rangle$,
which contains the contributions from the dynamical and the non-dynamical components.
As shown in Eqs.~\eqref{eq: t3 from three t1} and \eqref{eq: psi3 from t3},
$\hat{\psi}_{3,t_{3,t_1^3}}^R$ is sourced by $\hat{t}_{3,t_1^3}^R$,
and this source for GWs $\hat{t}_{3,t_1^3}^R$ is produced by the three different contributions from $S^{t^4} = S_4^{t^4} + S_{\mathrm{nd, s}}^{t^4} + S_{\mathrm{nd, v}}^{t^4}$.
In what follows, we will compute  these three contributions in order.

\subsubsection{diagram (a) from dynamical components}
\label{subsec: diagram a dynamical}
First, we consider the contribution from the dynamical components.
From Eqs.~\eqref{eq: t3 from three t1}, \eqref{eq: t3 source dynamical}, and \eqref{eq: psi3 from t3}, we obtain
\begin{align}
    &\left \langle
        \hat{\psi}_1^R(\tau, \bm{k}_1) \hat{\psi}_1^R(\tau, \bm{k}_2)
        \hat{\psi}_1^R(\tau, \bm{k}_3) \hat{\psi}_{3,t_{3,t_1^3}}^R(\tau, \bm{k}_4)
    \right \rangle_\mathrm{dynamical}
    \nonumber\\
    &=
    \int \mathrm{d}\eta_4 \, G_\psi(\tau, \eta_4, k_4) \mathcal{D}(\eta_4, k_4)
    \int \mathrm{d} \tilde{\eta}_4 \, G_t (\eta_4, \tilde{\eta}_4, k_4)
    c^{t^4} \int \frac{\mathrm{d}^3 p_4 \mathrm{d}^3 q_4 \mathrm{d}^3 r_4}{(2\pi)^6}
    \delta^{(3)}(-\bm{k}_4 + \bm{p}_4 + \bm{q}_4 + \bm{r}_4)
    \nonumber\\
    & \quad \times
    \left[
        e_{i j}^R(-\hat{\bm{k}}_4) e_{j k}^R(\hat{\bm{p}}_4)
        e_{k l}^R(\hat{\bm{q}}_4) e_{l i}^R(\hat{\bm{r}}_4)
        - e_{i j}^R(-\hat{\bm{k}}_4) e_{i j}^R(\hat{\bm{p}}_4) 
        e_{k l}^R(\hat{\bm{q}}_4) e_{k l}^R(\hat{\bm{r}}_4)
    \right]
    \nonumber\\
    & \quad \times
    \left \langle
        \hat{\psi}_1^R(\tau, \bm{k}_1) \hat{\psi}_1^R(\tau, \bm{k}_2)
        \hat{\psi}_1^R(\tau, \bm{k}_3) 
        \overline{
            \hat{t}_1^R(\tilde{\eta}_4, \bm{p}_4)
            \hat{t}_1^R(\tilde{\eta}_4, \bm{q}_4) \hat{t}_1^R(\tilde{\eta}_4, \bm{r}_4)
        }
    \right \rangle
    \nonumber\\
    &=
    (2\pi)^3 \delta^{(3)}(\bm{k}_1 + \bm{k}_2 + \bm{k}_3 + \bm{k}_4) 
    c^{t^4}
    \int \mathrm{d}\eta_4 \, G_\psi(\tau, \eta_4, k_4) \mathcal{D}(\eta_4, k_4)
    \int \mathrm{d} \tilde{\eta}_4 \, G_t (\eta_4, \tilde{\eta}_4, k_4)
    \nonumber\\
    & \quad \times
    \Psi_1^R(\tau, k_1) \Psi_1^R(\tau, k_2) \Psi_1^R(\tau, k_3)
    T_1^{R*}(\tilde{\eta}_4, k_1) T_1^{R*}(\tilde{\eta}_4, k_2)
    T_1^{R*}(\tilde{\eta}_4, k_3)
    \nonumber\\
    & \quad \times
    \left[
        \left(
            e_{i j}^R(-\hat{\bm{k}}_4) e_{j k}^R(-\hat{\bm{k}}_1)
            e_{k l}^R(-\hat{\bm{k}}_2) e_{l i}^R(-\hat{\bm{k}}_3)
            - 
            e_{i j}^R(-\hat{\bm{k}}_4) e_{i j}^R(-\hat{\bm{k}}_1) 
            e_{k l}^R(-\hat{\bm{k}}_2) e_{k l}^R(-\hat{\bm{k}}_3)
        \right)
    + (\mathrm{permutations\,of\,}\bm{k}_1,\,\bm{k}_2,\,\bm{k}_3)
    \right]
    \nonumber\\
    &=
    (2\pi)^3 \delta^{(3)}(\bm{k}_1 + \bm{k}_2 + \bm{k}_3 + \bm{k}_4) 
    c^{t^4}
    \int \mathrm{d}\eta_4 \, G_\psi(\tau, \eta_4, k_4) \mathcal{D}(\eta_4, k_4)
    \int \mathrm{d} \tilde{\eta}_4 \, G_t (\eta_4, \tilde{\eta}_4, k_4)
    \nonumber\\
    & \quad \times
    \Psi_1^R(\tau, k_1) \Psi_1^R(\tau, k_2) \Psi_1^R(\tau, k_3)
    T_1^{R*}(\tilde{\eta}_4, k_1) T_1^{R*}(\tilde{\eta}_4, k_2)
    T_1^{R*}(\tilde{\eta}_4, k_3)
    \nonumber\\  
    & \quad \times
    \left[
        \left(
            I_{-4,-1,-2,-3}^{t^4,1}
            - 
            I_{-4,-1,-2,-3}^{t^4,2}
        \right)
    + (\mathrm{permutations\,of\,}\bm{k}_1,\,\bm{k}_2,\,\bm{k}_3)
    \right]
    \nonumber\\
    &\equiv
    (2 \pi)^3 \delta^{(3)}(\bm{k}_1 + \bm{k}_2 + \bm{k}_3 + \bm{k}_4)
    F_\mathrm{d}^{(a)}(\bm{k}_1, \bm{k}_2, \bm{k}_3, \bm{k}_4),
\end{align}
where we define
\begin{align}
    I_{a,b,c,d}^{t^4, 1}
    &\equiv
    e_{i j}^R(\hat{\bm{k}}_a) e_{j k}^R(\hat{\bm{k}}_b)
    e_{k l}^R(\hat{\bm{k}}_c) e_{l i}^R(\hat{\bm{k}}_d)
    \\
    I_{a,b,c,d}^{t^4, 2}
    &\equiv
    e_{i j}^R(\hat{\bm{k}}_a) e_{i j}^R(\hat{\bm{k}}_b)
    e_{k l}^R(\hat{\bm{k}}_c) e_{k l}^R(\hat{\bm{k}}_d)    
\end{align}

For the permutation of $\hat{\psi}_1^R$ and $\hat{\psi}_{3, t_{3,t_1^3}}^R$, we obtain
\begin{align}
    &\left \langle
        \hat{\psi}_{3,t_{3,t_1^3}}^R(\tau, \bm{k}_1) \hat{\psi}_1^R(\tau, \bm{k}_2)
        \hat{\psi}_1^R(\tau, \bm{k}_3) \hat{\psi}_1^R(\tau, \bm{k}_4) 
    \right \rangle_\mathrm{dynamical}
    \nonumber\\
    & =
    (2 \pi)^3 \delta^{(3)}(\bm{k}_1 + \bm{k}_2 + \bm{k}_3 + \bm{k}_4)
    \tilde{F}_\mathrm{d}^{(a)}(\bm{k}_1, \bm{k}_2, \bm{k}_3, \bm{k}_4),
    \\
    &\left \langle
        \hat{\psi}_1^R(\tau, \bm{k}_2) \hat{\psi}_1^R(\tau, \bm{k}_2)
        \hat{\psi}_{3,t_{3,t_1^3}}^R(\tau, \bm{k}_3) \hat{\psi}_1^R(\tau, \bm{k}_4) 
    \right \rangle_\mathrm{dynamical}
    \nonumber\\
    & =
    (2\pi)^3 \delta^{(3)}(\bm{k}_1 + \bm{k}_2 + \bm{k}_3 + \bm{k}_4) 
    c^{t^4}
    \int \mathrm{d}\eta_3 \, G_\psi(\tau, \eta_3, k_3) \mathcal{D}(\eta_3, k_3)
    \int \mathrm{d} \tilde{\eta}_3 \, G_t (\eta_3, \tilde{\eta}_3, k_3)
    \nonumber\\
    & \quad \times
    \Psi_1^{R}(\tau, k_1) \Psi_1^{R}(\tau, k_2) T_1^{R}(\tilde{\eta}_3, k_4) 
    T_1^{R*}(\tilde{\eta}_3, k_1) T_1^{R*}(\tilde{\eta}_3, k_2) \Psi_1^{R*}(\tau, k_4)
    \nonumber\\
    & \quad \times
    \frac{1}{3}
    \left[
        \left(
            I_{-3,-4,-1,-2}^{t^4,1}
            - 
            I_{-3,-4,-1,-2}^{t^4,2}
        \right)
    + (\mathrm{permutations\,of\,}\bm{k}_1,\,\bm{k}_2,\,\bm{k}_4)
    \right]
    \nonumber\\
    &\equiv
    (2 \pi)^3 \delta^{(3)}(\bm{k}_1 + \bm{k}_2 + \bm{k}_3 + \bm{k}_4)
    G_\mathrm{d}^{(a)}(\bm{k}_1, \bm{k}_2, \bm{k}_3, \bm{k}_4),
    \\
    &\left \langle
        \ \hat{\psi}_1^R(\tau, \bm{k}_1) \hat{\psi}_{3,t_{3,t_1^3}}^R(\tau, \bm{k}_2)
        \hat{\psi}_1^R(\tau, \bm{k}_3) \hat{\psi}_1^R(\tau, \bm{k}_4) 
    \right \rangle_\mathrm{dynamical}
    \nonumber\\
    & =
    (2 \pi)^3 \delta^{(3)}(\bm{k}_1 + \bm{k}_2 + \bm{k}_3 + \bm{k}_4)
    \tilde{G}_\mathrm{d}^{(a)}(\bm{k}_1, \bm{k}_2, \bm{k}_3, \bm{k}_4),
\end{align}
where $\tilde{F}^{(a)}(\bm{k}_1, \bm{k}_2, \bm{k}_3, \bm{k}_4)$ is the complex conjugate of $F^{(a)}(\bm{k}_4, \bm{k}_3, \bm{k}_2, \bm{k}_1)$ except for $I^{t^4}$ factors, and the other $\tilde{F}$ and $\tilde{G}$ mentioned below are defined in the same way.

\subsubsection{diagram (a) from non-dynamical scalar components}
\label{subsec: diagram a non-dynamical scalar}
Next, we consider the contribution from the non-dynamical scalar components.
From Eqs.~\eqref{eq: t3 from three t1}, \eqref{eq: t3 source scalar}, and \eqref{eq: psi3 from t3}, we obtain
\begin{align}
    &\left \langle
        \hat{\psi}_1^R(\tau, \bm{k}_1) \hat{\psi}_1^R(\tau, \bm{k}_2)
        \hat{\psi}_1^R(\tau, \bm{k}_3) \hat{\psi}_{3,t_{3,t_1^3}}^R(\tau, \bm{k}_4)
    \right \rangle_\mathrm{scalar}
    \nonumber\\
    &=
    -\frac{1}{2}(2\pi)^3 \delta^{(3)}(\bm{k}_1 + \bm{k}_2 + \bm{k}_3 + \bm{k}_4) 
    \int \mathrm{d}\eta_4 \, G_\psi(\tau, \eta_4, k_4) \mathcal{D}(\eta_4, k_4)
    \int \mathrm{d} \tilde{\eta}_4 \, G_t (\eta_4, \tilde{\eta}_4, k_4)
    \Psi_1^R(\tau, k_1) \Psi_1^R(\tau, k_2) \Psi_1^R(\tau, k_3)
    \nonumber\\
    & \quad \times
    A_{\mathrm{s},-1,-4} A_{\mathrm{s},-2,-3}
    \nonumber\\
    & \quad \times
    \left[
        \left\{
            F_{\mathrm{s},1+4}(\tilde{\eta}_4)
            T_1^{R* \prime}(\tilde{\eta}_4, k_1)
            T_1^{R* \prime}(\tilde{\eta}_4, k_2)
            T_1^{R*}(\tilde{\eta}_4, k_3)
        \right.
    \right.
    \nonumber\\
    & \qquad
    \left.
        \left.
            + 
            \left[
                F_{\mathrm{s},1+4}(\tilde{\eta}_4)
                T_1^{R*}(\tilde{\eta}_4, k_1)
                T_1^{R* \prime}(\tilde{\eta}_4, k_2)
                T_1^{R*}(\tilde{\eta}_4, k_3)
            \right]'
        \right\}
        + (\mathrm{permutations\,of\,}\bm{k}_1,\,\bm{k}_2,\,\bm{k}_3)
    \right]
    \nonumber\\
    &\equiv
    (2 \pi)^3 \delta^{(3)}(\bm{k}_1 + \bm{k}_2 + \bm{k}_3 + \bm{k}_4)
    F_\mathrm{s}^{(a)}(\bm{k}_1, \bm{k}_2, \bm{k}_3, \bm{k}_4),
    \label{eq: Fs(a)}
    \\
    &\left \langle
        \hat{\psi}_{3,t_{3,t_1^3}}^R(\tau, \bm{k}_1) \hat{\psi}_1^R(\tau, \bm{k}_2)
        \hat{\psi}_1^R(\tau, \bm{k}_3) \hat{\psi}_1^R(\tau, \bm{k}_4) 
    \right \rangle_\mathrm{scalar}
    \nonumber\\
    &\equiv
    (2 \pi)^3 \delta^{(3)}(\bm{k}_1 + \bm{k}_2 + \bm{k}_3 + \bm{k}_4)
    \tilde{F}_\mathrm{s}^{(a)}(\bm{k}_1, \bm{k}_2, \bm{k}_3, \bm{k}_4),
    \\
    &\left \langle
        \hat{\psi}_1^R(\tau, \bm{k}_1) \hat{\psi}_1^R(\tau, \bm{k}_2)
        \hat{\psi}_{3,t_{3,t_1^3}}^R(\tau, \bm{k}_3) \hat{\psi}_1^R(\tau, \bm{k}_4) 
    \right \rangle_\mathrm{scalar}
    \nonumber\\
    &=
    -\frac{1}{6}(2\pi)^3 \delta^{(3)}(\bm{k}_1 + \bm{k}_2 + \bm{k}_3 + \bm{k}_4) 
    \int \mathrm{d}\eta_3 \, G_\psi(\tau, \eta_3, k_3) \mathcal{D}(\eta_3, k_3)
    \int \mathrm{d} \tilde{\eta}_3 \, G_t (\eta_3, \tilde{\eta}_3, k_3)
    \Psi_1^{R}(\tau, k_1) \Psi_1^{R}(\tau, k_2) \Psi_1^{R*}(\tau, k_4)
    \nonumber\\
    & \quad \times
    \left[
        A_{\mathrm{s},-4,-3} A_{\mathrm{s},-1,-2}
    \right.
    \nonumber\\
    & \qquad \times
    \left\{
        F_{\mathrm{s},3+4}(\tilde{\eta}_3)
        T_1^{R \prime}(\tilde{\eta}_3, k_4)
        T_1^{R* \prime}(\tilde{\eta}_3, k_1)
        T_1^{R*}(\tilde{\eta}_3, k_2)
    \right.
    \nonumber\\
    & \qquad \quad
    \left.
        +
        \left[
            F_{\mathrm{s},3+4}(\tilde{\eta}_3)
            T_1^{R}(\tilde{\eta}_3, k_4)
            T_1^{R* \prime}(\tilde{\eta}_3, k_1)
            T_1^{R*}(\tilde{\eta}_3, k_2)
        \right]'
    \right\}
    \nonumber\\
    & \qquad +
    \left.
        (\mathrm{permutations\,of\,}\bm{k}_1,\,\bm{k}_2,\,\bm{k}_4\,\mathrm{with\,replacements\,of}\,T_1^R\,\mathrm{and}\,T_1^{R*} )
    \right]
    \label{eq: Gs(a)}
    \nonumber\\
    &\equiv
    (2 \pi)^3 \delta^{(3)}(\bm{k}_1 + \bm{k}_2 + \bm{k}_3 + \bm{k}_4)
    G_\mathrm{s}^{(a)}(\bm{k}_1, \bm{k}_2, \bm{k}_3, \bm{k}_4),
    \\
    &\left \langle
        \hat{\psi}_1^R(\tau, \bm{k}_1) \hat{\psi}_{3,t_{3,t_1^3}}^R(\tau, \bm{k}_2) 
        \hat{\psi}_1^R(\tau, \bm{k}_3) \hat{\psi}_1^R(\tau, \bm{k}_4) 
    \right \rangle_\mathrm{scalar}
    \nonumber\\
    &\equiv
    (2 \pi)^3 \delta^{(3)}(\bm{k}_1 + \bm{k}_2 + \bm{k}_3 + \bm{k}_4)
    \tilde{G}_\mathrm{s}^{(a)}(\bm{k}_1, \bm{k}_2, \bm{k}_3, \bm{k}_4),
\end{align}
where ``permutations of $\bm{k}_1$, $\bm{k}_2$, $\bm{k}_4$ with replacements of $T_1^R$ and $T_1^{R*}$'' in Eq.~\eqref{eq: Gs(a)} represents the terms with permutations of $\bm{k}_1$, $\bm{k}_2$, $\bm{k}_4$ where $*$s are applied to $T_1^R$ with the arguments of $\bm{k}_1$ and $\bm{k}_2$.
Note that all terms with $A_\mathrm{s}$ vanish in the equilateral limit, $k_1 = k_2 = k_3 = k_4 = k$.

\subsubsection{diagram (a) from non-dynamical vector components}
\label{subsec: diagram a non-dynamical vector}

Finally, we consider the contribution from the non-dynamical vector components.
From Eqs.~\eqref{eq: t3 from three t1}, \eqref{eq: t3 source vector}, and \eqref{eq: psi3 from t3}, we obtain
\begin{align}
    &\left \langle
        \hat{\psi}_1^R(\tau, \bm{k}_1) \hat{\psi}_1^R(\tau, \bm{k}_2)
        \hat{\psi}_1^R(\tau, \bm{k}_3) \hat{\psi}_{3,t_{3,t_1^3}}^R(\tau, \bm{k}_4)
    \right \rangle_\mathrm{vector}
    \nonumber\\
    &=
    -\frac{1}{2}(2\pi)^3 \delta^{(3)}(\bm{k}_1 + \bm{k}_2 + \bm{k}_3 + \bm{k}_4) 
    \int \mathrm{d}\eta_4 \, G_\psi(\tau, \eta_4, k_4) \mathcal{D}(\eta_4, k_4)
    \int \mathrm{d} \tilde{\eta}_4 \, G_t (\eta_4, \tilde{\eta}_4, k_4)
    \Psi_1^R(\tau, k_1) \Psi_1^R(\tau, k_2) \Psi_1^R(\tau, k_3)
    \sum_{\lambda = R,L}
    \nonumber\\
    & \quad \times
    \left[
        A^\lambda_{\mathrm{v},-4,-1}
        A^\lambda_{\mathrm{v},-2,-3}
    \right.
    \nonumber\\
    & \qquad
    \times
    \left(
        F^\lambda_\mathrm{v,1+4}(\tilde{\eta}_4)
        T_1^{R* \prime}(\tilde{\eta}_4, k_1)
        T_1^{R*}(\tilde{\eta}_4,k_2) T_1^{R* \prime}(\tilde{\eta}_4,k_3)
        +
        \left\{
            F^\lambda_\mathrm{v,1+4}(\tilde{\eta}_4)
            T_1^{R*}(\tilde{\eta}_4, k_1)
            T_1^{R*}(\tilde{\eta}_4,k_2) T_1^{R* \prime}(\tilde{\eta}_4,k_3)
        \right\}'
    \right)
    \nonumber\\
    & \qquad
    \left.
        + (\mathrm{permutations\,of\,}\bm{k}_1,\,\bm{k}_2,\,\bm{k}_3)
    \right]
    \nonumber\\
    &\equiv
    (2 \pi)^3 \delta^{(3)}(\bm{k}_1 + \bm{k}_2 + \bm{k}_3 + \bm{k}_4)
    F_\mathrm{v}^{(a)}(\bm{k}_1, \bm{k}_2, \bm{k}_3, \bm{k}_4),
    \label{eq: Fv(a)}
    \\
    &\left \langle
        \hat{\psi}_{3,t_{3,t_1^3}}^R(\tau, \bm{k}_1) \hat{\psi}_1^R(\tau, \bm{k}_2)
        \hat{\psi}_1^R(\tau, \bm{k}_3) \hat{\psi}_1^R(\tau, \bm{k}_4) 
    \right \rangle_\mathrm{vector}
    \nonumber\\
    &\equiv
    (2 \pi)^3 \delta^{(3)}(\bm{k}_1 + \bm{k}_2 + \bm{k}_3 + \bm{k}_4)
    \tilde{F}_\mathrm{v}^{(a)}(\bm{k}_1, \bm{k}_2, \bm{k}_3, \bm{k}_4),
    \\
    &\left \langle
        \hat{\psi}_1^R(\tau, \bm{k}_1) \hat{\psi}_1^R(\tau, \bm{k}_2)
        \hat{\psi}_{3,t_{3,t_1^3}}^R(\tau, \bm{k}_3) \hat{\psi}_1^R(\tau, \bm{k}_4) 
    \right \rangle_\mathrm{vector}
    \nonumber\\
    &=
    -\frac{1}{6}(2\pi)^3 \delta^{(3)}(\bm{k}_1 + \bm{k}_2 + \bm{k}_3 + \bm{k}_4) 
    \int \mathrm{d}\eta_3 \, G_\psi(\tau, \eta_3, k_3) \mathcal{D}(\eta_3, k_3)
    \int \mathrm{d} \tilde{\eta}_3 \, G_t (\eta_3, \tilde{\eta}_3, k_3)
    \Psi_1^R(\tau, k_1) \Psi_1^R(\tau, k_2) \Psi_1^{R*}(\tau, k_4)
    \sum_{\lambda = R,L}
    \nonumber\\
    & \quad \times
    \left[
        A^\lambda_{\mathrm{v},-3,-4}
        A^\lambda_{\mathrm{v},-1,-2}
    \right.
    \nonumber\\
    & \qquad
    \times
    \left(
        F^\lambda_\mathrm{v,3+4}(\tilde{\eta}_3)
        T_1^{R \prime}(\tilde{\eta}_3, k_4)
        T_1^{R*}(\tilde{\eta}_3,k_1) T_1^{R* \prime}(\tilde{\eta}_3,k_2)
        +
        \left\{
            F^\lambda_\mathrm{v,3+4}(\tilde{\eta}_3)
            T_1^{R}(\tilde{\eta}_3, k_4)
            T_1^{R*}(\tilde{\eta}_3,k_1) T_1^{R* \prime}(\tilde{\eta}_3,k_2)
        \right\}'
    \right)
    \nonumber\\
    & \qquad +
    \left.
        (\mathrm{permutations\,of\,}\bm{k}_1,\,\bm{k}_2,\,\bm{k}_4\,\mathrm{with\,replacements\,of}\,T_1^R\,\mathrm{and}\,T_1^{R*} )
    \right]
    \label{eq: Gv(a)}
    \nonumber\\
    &\equiv
    (2 \pi)^3 \delta^{(3)}(\bm{k}_1 + \bm{k}_2 + \bm{k}_3 + \bm{k}_4)
    G_\mathrm{v}^{(a)}(\bm{k}_1, \bm{k}_2, \bm{k}_3, \bm{k}_4),
    \\
    &\left \langle
        \hat{\psi}_1^R(\tau, \bm{k}_1) \hat{\psi}_{3,t_{3,t_1^3}}^R(\tau, \bm{k}_2) 
        \hat{\psi}_1^R(\tau, \bm{k}_3) \hat{\psi}_1^R(\tau, \bm{k}_4) 
    \right \rangle_\mathrm{vector}
    \nonumber\\
    &\equiv
    (2 \pi)^3 \delta^{(3)}(\bm{k}_1 + \bm{k}_2 + \bm{k}_3 + \bm{k}_4)
    \tilde{G}_\mathrm{v}^{(a)}(\bm{k}_1, \bm{k}_2, \bm{k}_3, \bm{k}_4),
\end{align}
where ``permutations of $\bm{k}_1$, $\bm{k}_2$, $\bm{k}_4$ with replacements of $T_1^R$ and $T_1^{R*}$'' in Eq.~\eqref{eq: Gv(a)} is defined in the same way as in Eq.~\eqref{eq: Gs(a)}.

\subsection{diagram (b)}
\label{subsec: diagram b}

Here we consider the diagram (b): 
$\left \langle \hat{\psi}_1^R(\tau, \bm{k}_1) \hat{\psi}_1^R(\tau, \bm{k}_2)  \hat{\psi}_1^R(\tau, \bm{k}_3) \hat{\psi}_{3,t_1^3}^R(\tau, \bm{k}_4)\right \rangle$.
In the following, we obtain the trispectra by substituting the source term, which is divided into three parts as 
$S^{\psi t^3} = S_4^{\psi t^3} + S_{\mathrm{nd, s}}^{\psi t^3} + S_{\mathrm{nd, v}}^{\psi t^3}$ into $\hat{\psi}_{3,t_1^3}$ in Eq.~\eqref{eq: psi3 from three t1}.
\subsubsection{diagram (b) from dynamical components}
\label{subsec: diagram b dynamical}

First, we consider the contribution from the dynamical components.
By substituting Eq.~\eqref{eq: psi3 source dynamical} into Eq.~\eqref{eq: psi3 from three t1}, we obtain
\begin{align}
    &\left \langle
        \hat{\psi}_1^R(\tau, \bm{k}_1) \hat{\psi}_1^R(\tau, \bm{k}_2)
        \hat{\psi}_1^R(\tau, \bm{k}_3) \hat{\psi}_{3,t_1^3}^R(\tau, \bm{k}_4)
    \right \rangle_\mathrm{dynamical}
    \nonumber\\
    &=
    (2\pi)^3 \delta^{(3)}(\bm{k}_1 + \bm{k}_2 + \bm{k}_3 + \bm{k}_4) 
    c^{\psi t^3}
    \int \mathrm{d}\eta_4 \, G_\psi(\tau, \eta_4, k_4)
    \Psi_1^R(\tau, k_1) \Psi_1^R(\tau, k_2) \Psi_1^R(\tau, k_3)
    T_1^{R*}(\eta_4, k_1) T_1^{R*}(\eta_4, k_2)
    T_1^{R*}(\eta_4, k_3)
    \nonumber\\  
    & \quad \times
    \left[
        I^{\psi t^3}_{-4,-1,-2,-3}
        + (\mathrm{permutations\,of\,}\bm{k}_1,\,\bm{k}_2,\,\bm{k}_3)
    \right]
    \nonumber\\
    &\equiv
    (2 \pi)^3 \delta^{(3)}(\bm{k}_1 + \bm{k}_2 + \bm{k}_3 + \bm{k}_4)
    F_\mathrm{d}^{(b)}(\bm{k}_1, \bm{k}_2, \bm{k}_3, \bm{k}_4),
\end{align}
where we define
\begin{align}
    I_{a,b,c,d}^{\psi t^3}(\tau)
    &\equiv
    I^{\psi t^3}(\tau,\bm{k}_a,\bm{k}_b,\bm{k}_c,\bm{k}_d)
    \nonumber\\
    &=
    e_{i j}^R(\hat{\bm{k}}_a)
    \left[
        e_{i j}^R(\hat{\bm{k}}_b) e_{k l}^R(\hat{\bm{k}}_c) e_{k l}^R(\hat{\bm{k}}_d)
        -2 e_{j k}^R(\hat{\bm{k}}_b) e_{k l}^R(\hat{\bm{k}}_c) e_{l i}^R(\hat{\bm{k}}_d)
        + \frac{\tau}{m_Q} \epsilon^{l m n} i 
        \left(
            k_b^k e_{l i}^R(\hat{\bm{k}}_b) - k_b^i e_{l k}^R(\hat{\bm{k}}_b)
        \right)
        e_{m k}^R(\hat{\bm{k}}_c) e_{n j}^R(\hat{\bm{k}}_d)
    \right]
\end{align}

For the permutation of $\hat{\psi}_1^R$ and $\hat{\psi}_{3, t_1^3}^R$,
we obtain
\begin{align}
    &\left \langle
        \hat{\psi}_{3, t_1^3}^R(\tau, \bm{k}_1) \hat{\psi}_1^R(\tau, \bm{k}_2)
        \hat{\psi}_1^R(\tau, \bm{k}_3) \hat{\psi}_1^R(\tau, \bm{k}_4) 
    \right \rangle_\mathrm{dynamical}
    \nonumber\\
    & =
    (2 \pi)^3 \delta^{(3)}(\bm{k}_1 + \bm{k}_2 + \bm{k}_3 + \bm{k}_4)
    \tilde{F}_\mathrm{d}^{(b)}(\bm{k}_1, \bm{k}_2, \bm{k}_3, \bm{k}_4),
    \\
    &\left \langle
        \hat{\psi}_1^R(\tau, \bm{k}_2) \hat{\psi}_1^R(\tau, \bm{k}_2)
        \hat{\psi}_{3, t_1^3}^R(\tau, \bm{k}_3) \hat{\psi}_1^R(\tau, \bm{k}_4) 
    \right \rangle_\mathrm{dynamical}
    \nonumber\\
    & =
    (2\pi)^3 \delta^{(3)}(\bm{k}_1 + \bm{k}_2 + \bm{k}_3 + \bm{k}_4) 
    c^{\psi t^3}
    \int \mathrm{d}\eta_3 \, G_\psi(\tau, \eta_3, k_3)
    \Psi_1^{R}(\tau, k_1) \Psi_1^{R}(\tau, k_2) T_1^{R}(\eta_3, k_4) 
    T_1^{R*}(\eta_3, k_1) T_1^{R*}(\eta_3, k_2) \Psi_1^{R*}(\tau, k_4)
    \nonumber\\
    & \quad \times
    \frac{1}{3}
    \left[
        I_{-3,-4,-1,-2}^{\psi t^3}
        + (\mathrm{permutations\,of\,}\bm{k}_1,\,\bm{k}_2,\,\bm{k}_4)
    \right]
    \nonumber\\
    &\equiv
    (2 \pi)^3 \delta^{(3)}(\bm{k}_1 + \bm{k}_2 + \bm{k}_3 + \bm{k}_4)
    G_\mathrm{d}^{(b)}(\bm{k}_1, \bm{k}_2, \bm{k}_3, \bm{k}_4),
    \\
    &\left \langle
        \ \hat{\psi}_1^R(\tau, \bm{k}_1) \hat{\psi}_{3, t_1^3}^R(\tau, \bm{k}_2)
        \hat{\psi}_1^R(\tau, \bm{k}_3) \hat{\psi}_1^R(\tau, \bm{k}_4) 
    \right \rangle_\mathrm{dynamical}
    \nonumber\\
    & =
    (2 \pi)^3 \delta^{(3)}(\bm{k}_1 + \bm{k}_2 + \bm{k}_3 + \bm{k}_4)
    \tilde{G}_\mathrm{d}^{(b)}(\bm{k}_1, \bm{k}_2, \bm{k}_3, \bm{k}_4).
\end{align}
\subsubsection{diagram (b) from non-dynamical scalar components}
\label{subsec: diagram b non-dynamical scalar}

Next, we consider the contribution from the non-dynamical scalar components.
By substituting Eq.~\eqref{eq: psi3 source scalar} into Eq.~\eqref{eq: psi3 from three t1}, we obtain
\begin{align}
    &\left \langle
        \hat{\psi}_1^R(\tau, \bm{k}_1) \hat{\psi}_1^R(\tau, \bm{k}_2)
        \hat{\psi}_1^R(\tau, \bm{k}_3) \hat{\psi}_{3,t_1^3}^R(\tau, \bm{k}_4)
    \right \rangle_\mathrm{scalar}
    \nonumber\\
    &=
    -\frac{1}{2}(2\pi)^3 \delta^{(3)}(\bm{k}_1 + \bm{k}_2 + \bm{k}_3 + \bm{k}_4) 
    \int \mathrm{d}\eta_4 \, G_\psi(\tau, \eta_4, k_4)
    \Psi_1^R(\tau, k_1) \Psi_1^R(\tau, k_2) \Psi_1^R(\tau, k_3)
    T_1^{R*}(\eta_4, k_2) T_1^{R*}(\eta_4, k_3)
    \nonumber\\
    & \quad \times
    F_{\mathrm{s},1+4}(\eta_4)
    A_{\mathrm{s},-2,-3}
    \nonumber\\
    & \quad \times
    \left[
    \left(
        \frac{\eta_4}{m_Q} T_1^{R* \prime}(\eta_4,k_2)
        S_{\mathrm{s},-1,-4}
        -
        \frac{c^{\psi t^2}}{c^{t^3}}
        \left(
            T_1^{R* \prime}(\eta_4,k_2) 
            +
            \frac{T_1^{R*}(\eta_4,k_2)}{\tau}
        \right)
        A_{\mathrm{s},-1,-4}
    \right)
        + (\mathrm{permutations\,of\,}\bm{k}_1,\,\bm{k}_2,\,\bm{k}_3)
    \right]
    \nonumber\\
    &\equiv
    (2 \pi)^3 \delta^{(3)}(\bm{k}_1 + \bm{k}_2 + \bm{k}_3 + \bm{k}_4)
    F_\mathrm{s}^{(b)}(\bm{k}_1, \bm{k}_2, \bm{k}_3, \bm{k}_4),
    \\
    &\left \langle
        \hat{\psi}_{3,t_1^3}^R(\tau, \bm{k}_1) \hat{\psi}_1^R(\tau, \bm{k}_2)
        \hat{\psi}_1^R(\tau, \bm{k}_3) \hat{\psi}_1^R(\tau, \bm{k}_4) 
    \right \rangle_\mathrm{scalar}
    \nonumber\\
    &\equiv
    (2 \pi)^3 \delta^{(3)}(\bm{k}_1 + \bm{k}_2 + \bm{k}_3 + \bm{k}_4)
    \tilde{F}_\mathrm{s}^{(b)}(\bm{k}_1, \bm{k}_2, \bm{k}_3, \bm{k}_4),
    \\
    &\left \langle
        \hat{\psi}_1^R(\tau, \bm{k}_1) \hat{\psi}_1^R(\tau, \bm{k}_2)
        \hat{\psi}_{3,t_1^3}^R(\tau, \bm{k}_3) \hat{\psi}_1^R(\tau, \bm{k}_4) 
    \right \rangle_\mathrm{scalar}
    \nonumber\\
    &=
    -\frac{1}{6}(2\pi)^3 \delta^{(3)}(\bm{k}_1 + \bm{k}_2 + \bm{k}_3 + \bm{k}_4) 
    \int \mathrm{d}\eta_3 \, G_\psi(\tau, \eta_3, k_3)
    \Psi_1^{R}(\tau, k_1) \Psi_1^{R}(\tau, k_2) \Psi_1^{R*}(\tau, k_4)
    \nonumber\\
    & \quad \times
    \left[
        A_{\mathrm{s},-1,-2}
        T_1^{R*}(\eta_3, k_1) T_1^{R*}(\eta_3, k_2)
    \right.
    \nonumber\\
    & \qquad \times
    \left(
        \frac{\tau}{m_Q} T_1^{R \prime}(\eta_3,k_4)
        S_{\mathrm{s},-4,-3}
        -
        \frac{c^{\psi t^2}}{c^{t^3}}
        \left(
            T_1^{R \prime}(\eta_3,k_4) 
            +
            \frac{T_1^{R}(\eta_3,k_4)}{\eta_3}
        \right)
        A_{\mathrm{s},-4,-3}
    \right)
    \nonumber\\
    & \qquad +
    \left.
        (\mathrm{permutations\,of\,}\bm{k}_1,\,\bm{k}_2,\,\bm{k}_4\,\mathrm{with\,replacements\,of}\,T_1^R\,\mathrm{and}\,T_1^{R*} )
    \right]
    \label{eq: Gs(b)}
    \nonumber\\
    &\equiv
    (2 \pi)^3 \delta^{(3)}(\bm{k}_1 + \bm{k}_2 + \bm{k}_3 + \bm{k}_4)
    G_\mathrm{s}^{(a)}(\bm{k}_1, \bm{k}_2, \bm{k}_3, \bm{k}_4),
    \\
    &\left \langle
        \hat{\psi}_1^R(\tau, \bm{k}_1) \hat{\psi}_{3,t_1^3}^R(\tau, \bm{k}_2) 
        \hat{\psi}_1^R(\tau, \bm{k}_3) \hat{\psi}_1^R(\tau, \bm{k}_4) 
    \right \rangle_\mathrm{scalar}
    \nonumber\\
    &\equiv
    (2 \pi)^3 \delta^{(3)}(\bm{k}_1 + \bm{k}_2 + \bm{k}_3 + \bm{k}_4)
    \tilde{G}_\mathrm{s}^{(b)}(\bm{k}_1, \bm{k}_2, \bm{k}_3, \bm{k}_4),
\end{align}
where ``permutations of $\bm{k}_1$, $\bm{k}_2$, $\bm{k}_4$ with replacements of $T_1^R$ and $T_1^{R*}$'' in Eq.~\eqref{eq: Gs(b)} is defined in the same way as in Eq.~\eqref{eq: Gs(a)}.


\subsubsection{diagram (b) from non-dynamical vector components}
\label{subsec: diagram b non-dynamical vector}

Finally, we consider the contribution from the non-dynamical vector components.
By substituting Eq.~\eqref{eq: psi3 source vector} into Eq.~\eqref{eq: psi3 from three t1}, we obtain
\begin{align}
    &\left \langle
        \hat{\psi}_1^R(\tau, \bm{k}_1) \hat{\psi}_1^R(\tau, \bm{k}_2)
        \hat{\psi}_1^R(\tau, \bm{k}_3) \hat{\psi}_{3,t_1^3}^R(\tau, \bm{k}_4)
    \right \rangle_\mathrm{vector}
    \nonumber\\
    &=
    -\frac{1}{2}(2\pi)^3 \delta^{(3)}(\bm{k}_1 + \bm{k}_2 + \bm{k}_3 + \bm{k}_4) 
    \int \mathrm{d}\eta_4 \, G_\psi(\tau, \eta_4, k_4)
    \Psi_1^R(\tau, k_1) \Psi_1^R(\tau, k_2) \Psi_1^R(\tau, k_3)
    \sum_{\lambda = R,L}
    \nonumber\\
    & \quad \times
    \left[
        F^\lambda_\mathrm{v,1+4}(\eta_4)
        \left(
            \frac{\tau}{m_Q} P_{\mathrm{v},-1,-4}
            T_1^{R* \prime}(\eta_4, k_1)
            -
            \frac{c^{\psi t^2}}{c^{t^3}}
            A_{\mathrm{v},-1,-4}
            \left[ 
                T_1^{R* \prime}(\eta_4, k_1) 
                +
                \frac{T_1^{R*}(\eta_4, k_1)}{\eta_4}
            \right]
        \right)
    \right.
    \nonumber\\
    & \qquad \quad \times
    A_{\mathrm{v},-2,-3}
    T_1^{R*}(\eta_4, k_2) T_1^{R* \prime}(\eta_4, k_3) 
    \nonumber\\
    & \qquad
    \left.
        + (\mathrm{permutations\,of\,}\bm{k}_1,\,\bm{k}_2,\,\bm{k}_3)
    \right]
    \nonumber\\
    &\equiv
    (2 \pi)^3 \delta^{(3)}(\bm{k}_1 + \bm{k}_2 + \bm{k}_3 + \bm{k}_4)
    F_\mathrm{v}^{(b)}(\bm{k}_1, \bm{k}_2, \bm{k}_3, \bm{k}_4),
    \\
    &\left \langle
        \hat{\psi}_{3,t_1^3}^R(\tau, \bm{k}_1) \hat{\psi}_1^R(\tau, \bm{k}_2)
        \hat{\psi}_1^R(\tau, \bm{k}_3) \hat{\psi}_1^R(\tau, \bm{k}_4) 
    \right \rangle_\mathrm{vector}
    \nonumber\\
    &\equiv
    (2 \pi)^3 \delta^{(3)}(\bm{k}_1 + \bm{k}_2 + \bm{k}_3 + \bm{k}_4)
    \tilde{F}_\mathrm{v}^{(b)}(\bm{k}_1, \bm{k}_2, \bm{k}_3, \bm{k}_4),
    \\
    &\left \langle
        \hat{\psi}_1^R(\tau, \bm{k}_1) \hat{\psi}_1^R(\tau, \bm{k}_2)
        \hat{\psi}_{3,t_{3,t_1^3}}^R(\tau, \bm{k}_3) \hat{\psi}_1^R(\tau, \bm{k}_4) 
    \right \rangle_\mathrm{vector}
    \nonumber\\
    &=
    -\frac{1}{6}(2\pi)^3 \delta^{(3)}(\bm{k}_1 + \bm{k}_2 + \bm{k}_3 + \bm{k}_4) 
    \int \mathrm{d}\eta_3 \, G_\psi(\tau, \eta_3, k_3)
    \Psi_1^R(\tau, k_1) \Psi_1^R(\tau, k_2) \Psi_1^{R*}(\tau, k_4)
    \sum_{\lambda = R,L}
    \nonumber\\
    & \quad \times
    \left[
        F^\lambda_\mathrm{v,3+4}(\eta_3)
        \left(
            \frac{\tau}{m_Q} P_{\mathrm{v},-4,-3}
            T_1^{R \prime}(\eta_3, k_4)
            -
            \frac{c^{\psi t^2}}{c^{t^3}}
            A_{\mathrm{v},-4,-3}
            \left[ 
                T_1^{R \prime}(\eta_3, k_4) 
                +
                \frac{T_1^{R}(\eta_3, k_4)}{\eta_3}
            \right]
        \right)
    \right.
    \nonumber\\
    & \qquad \quad \times
    A_{\mathrm{v},-1,-2}
    T_1^{R*}(\eta_3, k_1) T_1^{R* \prime}(\eta_3, k_2) 
    \nonumber\\
    & \qquad +
    \left.
        (\mathrm{permutations\,of\,}\bm{k}_1,\,\bm{k}_2,\,\bm{k}_4\,\mathrm{with\,replacements\,of}\,T_1^R\,\mathrm{and}\,T_1^{R*} )
    \right]
    \label{eq: Gv(b)}
    \nonumber\\
    &\equiv
    (2 \pi)^3 \delta^{(3)}(\bm{k}_1 + \bm{k}_2 + \bm{k}_3 + \bm{k}_4)
    G_\mathrm{v}^{(b)}(\bm{k}_1, \bm{k}_2, \bm{k}_3, \bm{k}_4),
    \\
    &\left \langle
        \hat{\psi}_1^R(\tau, \bm{k}_1) \hat{\psi}_{3,t_1^3}^R(\tau, \bm{k}_2) 
        \hat{\psi}_1^R(\tau, \bm{k}_3) \hat{\psi}_1^R(\tau, \bm{k}_4) 
    \right \rangle_\mathrm{vector}
    \nonumber\\
    &\equiv
    (2 \pi)^3 \delta^{(3)}(\bm{k}_1 + \bm{k}_2 + \bm{k}_3 + \bm{k}_4)
    \tilde{G}_\mathrm{v}^{(b)}(\bm{k}_1, \bm{k}_2, \bm{k}_3, \bm{k}_4),
\end{align}
where ``permutations of $\bm{k}_1$, $\bm{k}_2$, $\bm{k}_4$ with replacements of $T_1^R$ and $T_1^{R*}$'' in Eq.~\eqref{eq: Gv(b)} is defined in the same way as in Eq.~\eqref{eq: Gs(a)}.
\endgroup
\section{Evaluation of \texorpdfstring{$g_\mathrm{NL}$}{}-type GW trispectrum}
\label{sec: gNL evaluation}

In this section, we estimate the magnitude of the GW trispectra in the super-horizon limit.
A trispectrum depends on four momenta and
we need to fix six parameters to specify the momentum configuration even after exploiting the momentum conservation and the coordinate choice.
In the rest of this paper, however, we concentrate on the equilateral configuration, which involves only two parameters. The detail of the equilateral configuration is given in App.~\ref{app: polarization tensor}.
Since the linear tensor perturbation of the gauge field $\hat{t}_1(\tau,\bm k)$ is drastically amplified around the horizon crossing and then quickly decays afterwards, the induced non-linear perturbations have to carry a similar momentum to $k$. 
For example, when $\hat{t}_1(\bm k_1)$
and $\hat{t}_1(\bm k_2)$ produce $\hat{\psi}_2(\bm k_1+\bm k_2)$, in order for the both of $\hat{t}_1$ to have large amplitudes at the same time and significantly generate $\hat{\psi}_2$, their momenta should be approximately same, $k_1\simeq k_2$, and hence $|\bm k_1+\bm k_2|$ is also a similar value. The same is true for the cases of the three point vertices.
Therefore, it is naturally expected that the GW trispectrum has a significant signal only when four momenta have similar sizes. 
Indeed, it has been shown that the induced GW bispectrum peaks at around the equilateral configuration and it matches the equilateral shape of bispectrum by more than 90\%~\cite{Agrawal:2017awz,Agrawal:2018mrg}. 
For this reason, the trispectrum signal for the equilateral configuration is a main target in the following analysis.

\subsection{Evaluation of diagram (a)}
\label{subsec: evaluation of diagram a}

\subsubsection{dynamical contributions}
\label{subsubsec: evaluation of diagram a dynamical}

First we evaluate only the contribution from diagram (a) with the dynamical components.

\begin{figure}[htpb]
    \includegraphics[clip, width=0.8\columnwidth]{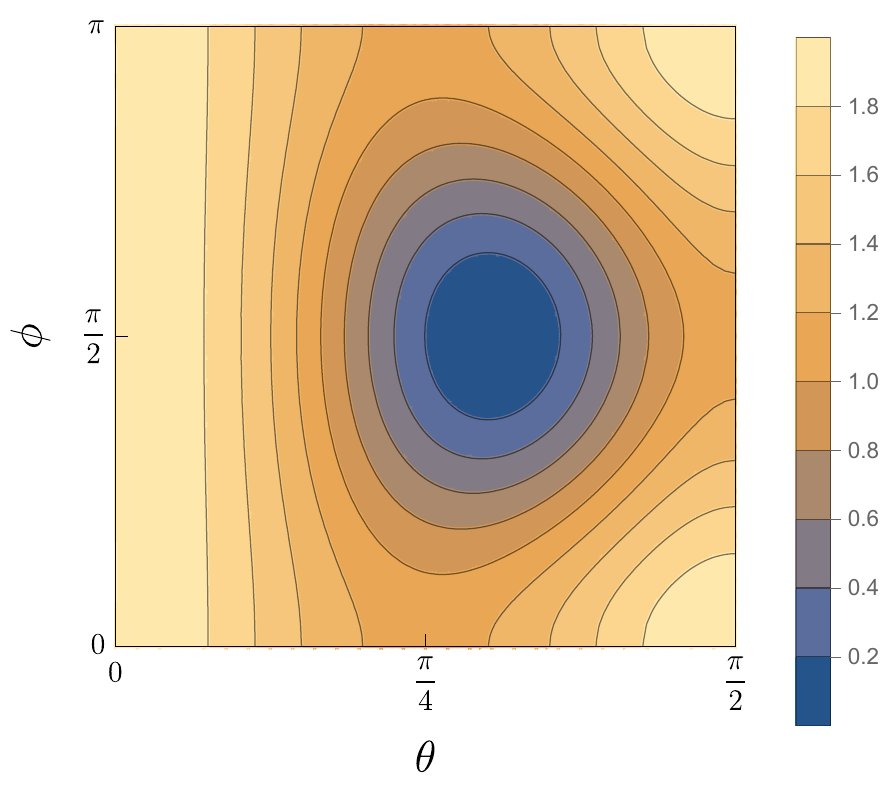}
    \caption{
    Evaluation of $|C^{(a)}|$ as a function of $\theta$ and $\phi$.
    }
    \label{fig: evaluation of C^(a)}
\end{figure}

Therefore,
\begin{align}
    &
    \frac{\left \langle
        \hat{\psi}^R(\tau, \bm{k}_1) \hat{\psi}^R(\tau, \bm{k}_2)
        \hat{\psi}^R(\tau, \bm{k}_3) \hat{\psi}^R(\tau, \bm{k}_4)
    \right \rangle
    }
    {(2 \pi)^3 \delta^{(3)}(\bm{k}_1 + \bm{k}_2 + \bm{k}_3 + \bm{k}_4)}
    \nonumber\\
    & \supset
    F_\mathrm{d}^{(a)}(\bm{k}_1, \bm{k}_2, \bm{k}_3, \bm{k}_4)
    \nonumber\\
    & =
    g^2
    \int \mathrm{d}\eta_4 \, G_\psi(\tau, \eta_4, k) \mathcal{D}(\eta_4, k)
    \int \mathrm{d} \tilde{\eta}_4 \, G_t (\eta_4, \tilde{\eta}_4, k)
    \nonumber\\
    & \quad \times
    \left[
        \Psi_1^R(\tau, k) T_1^{R*}(\tilde{\eta}_4, k)
    \right]^3
    \left[
        \left(
            I_{-4,-1,-2,-3}^{t^4,1}
            - 
            I_{-4,-1,-2,-3}^{t^4,2}
        \right)
    + (\mathrm{permutations\,of\,}\bm{k}_1,\,\bm{k}_2,\,\bm{k}_3)
    \right]
    \nonumber\\
    & =
    - g^2 C^{(a)}
    \int \mathrm{d}\eta \, G_\psi(\tau, \eta, k) \mathcal{D}(\eta, k)
    \int \mathrm{d} \tilde{\eta} \, G_t (\eta, \tilde{\eta}, k)
    \left[
        \Psi_1^R(\tau, k) T_1^{R*}(\tilde{\eta}, k)
    \right]^3
    \nonumber\\
    & =
    - g^2 C^{(a)}
    \left[
        \Psi_1^R(\tau, k)
    \right]^3
    \int_{-\infty}^{\tau} \mathrm{d}\eta \,
    \frac
    {k(\eta - \tau) \cos\left( k(\tau-\eta) \right)
    + (1 + k^2\tau \eta) \sin\left( k(\tau - \eta) \right)}
    {k^3 \tau \eta}
    \left[
        \frac{2\sqrt{\epsilon_B}}{m_Q \eta}\partial_\eta
        +\frac{2\sqrt{\epsilon_B}}{\eta^2}(m_Q + k\eta)
    \right]
    \nonumber\\
    & \quad \times
    \int_{-\infty}^{\eta} \mathrm{d} \tilde{\eta} \,
    \frac{1}{k}
    e^{\pi(2m_Q + m_Q^{-1})}
    \mathrm{Im} \left[
        W_{\beta, \alpha} (2i k\eta)^* W_{\beta, \alpha}(2i k\tilde{\eta})
    \right]
    \left[
        \frac{1}{\sqrt{2k}} e^{\pi(2m_Q + m_Q^{-1})/2}
        W_{\beta, \alpha}(2ik\tilde{\eta})^*
    \right]^3.
\end{align}
Here, we parametrize $I^{t^4}$ as
\begin{equation}
    \left( I^{t^4,1}_{-4,-1,-2,-3} - I^{t^4,2}_{-4,-1,-2,-3} \right)
    + (\mathrm{permutations\,of\,}\bm{k}_1,\,\bm{k}_2,\,\bm{k}_3)
    =
    C^{(a)}.
\end{equation}
$C(a)$ is a function of the angles of the equilateral configuration and evaluated as in Fig.~\ref{fig: evaluation of C^(a)}.
$|C^{(a)}|$ takes values from $0$ to $2$.

By taking the super-horizon limit $x \equiv - k\tau \to +0$, we obtain
\begin{align}
    &
    F_\mathrm{d}^{(a)}(\bm{k}_1, \bm{k}_2, \bm{k}_3, \bm{k}_4)
    \nonumber\\
    & =
    -4 \sqrt{\epsilon_B} g^2 C^{(a)}
    e^{5\pi(2m_Q + m_Q^{-1})/2}
    \left[
        \Psi_1^R(k\tau \to 0)
    \right]^3
    (2k)^{-5/2}
    \int_{-\infty}^{0} \mathrm{d}\eta \,
    \frac
    {k \eta \cos\left( k \eta \right)
    - \sin\left( k \eta \right)}
    {k^3 \tau \eta}
    \left[
        \frac{\partial_\eta}{m_Q \eta}
        +\frac{m_Q + k\eta}{\eta^2}
    \right]
    \nonumber\\
    & \quad \times
    \int_{-\infty}^{\eta} \mathrm{d} \tilde{\eta} \,
    \mathrm{Im} \left[
        W_{\beta, \alpha} (2i k\eta)^* W_{\beta, \alpha}(2i k\tilde{\eta})
    \right]
    \left[
        W_{\beta, \alpha}(2ik\tilde{\eta})^*
    \right]^3
    \nonumber\\
    & =
    -\frac{\sqrt{\epsilon_B} g^2 C^{(a)}}{2^{1/2}k^{9/2}\tau}
    e^{5\pi(2m_Q + m_Q^{-1})/2}
    \left[
        \frac{\sqrt{\epsilon_B}}{\sqrt{2k}k\tau} \mathcal{F}(m_Q)
    \right]^3
    \int_0^{x_\mathrm{max}} \frac{\mathrm{d}y}{y^2}
    (y \cos y - \sin y)
    \left[
        \frac{\partial_y}{m_Q}
        +\frac{m_Q}{y} -1
    \right]
    \nonumber\\
    & \quad \times
    \int_y^{x_\mathrm{max}} \mathrm{d} \tilde{y} \,
    \mathrm{Im} \left[
        W_{\beta, \alpha} (-2i y)^* W_{\beta, \alpha}(-2i \tilde{y})
    \right]
    \left[
        W_{\beta, \alpha}(-2i\tilde{y})^*
    \right]^3
    \nonumber\\
    & =
    -\frac{\epsilon_B^2 g^2 C^{(a)}}{2^2 k^9 \tau^4}
    e^{5\pi(2m_Q + m_Q^{-1})/2}
    \left[
         \mathcal{F}(m_Q)
    \right]^3
    \int_0^{x_\mathrm{max}} \frac{\mathrm{d}y}{y^2}
    (y \cos y - \sin y)
    \left[
        \frac{\partial_y}{m_Q}
        +\frac{m_Q}{y} -1
    \right]
    \nonumber\\
    & \quad \times
    \int_y^{x_\mathrm{max}} \mathrm{d} \tilde{y} \,
    \mathrm{Im} \left[
        W_{\beta, \alpha} (-2i y)^* W_{\beta, \alpha}(-2i \tilde{y})
    \right]
    \left[
        W_{\beta, \alpha}(-2i\tilde{y})^*
    \right]^3,
\end{align}
where we have introduced the UV cutoff $x_\mathrm{max} \equiv 2m_Q + m_Q^{-1} + \sqrt{2m_Q^2 + 2 + m_Q^{-2}}$, at which $T_1^R$ starts undergoing a tachyonic instability, to eliminate unphysical vacuum contributions.

In the same way, we evaluate $G_\mathrm{d}^{(a)}$ as
\begin{align}
    &
    G_\mathrm{d}^{(a)}(\bm{k}_1, \bm{k}_2, \bm{k}_3, \bm{k}_4)
    \nonumber\\
    & =
    -\frac{1}{3}
    \frac{\epsilon_B^2 g^2 C^{(a)}}{2^2 k^9 \tau^4}
    e^{5\pi(2m_Q + m_Q^{-1})/2}
    \left[
         \mathcal{F}(m_Q)
    \right]^2
    \mathcal{F}^*(m_Q)
    \int_0^{x_\mathrm{max}} \frac{\mathrm{d}y}{y^2}
    (y \cos y - \sin y)
    \left[
        \frac{\partial_y}{m_Q}
        +\frac{m_Q}{y} -1
    \right]
    \nonumber\\
    & \quad \times
    \int_y^{x_\mathrm{max}} \mathrm{d} \tilde{y} \,
    \mathrm{Im} \left[
        W_{\beta, \alpha} (-2i y)^* W_{\beta, \alpha}(-2i \tilde{y})
    \right]
    W_{\beta, \alpha}(-2i\tilde{y})
    \left[
        W_{\beta, \alpha}(-2i\tilde{y})^*
    \right]^2.
\end{align}

Since $\psi_{i j} = a M_\mathrm{P} h_{i j}/2$ and $\tau = -a H$,
\begin{align}
    T_h^{RRRR}
    & \supset
    \frac{4 \epsilon_B^2 g^2 C^{(a)}}{k^9}
    \left( \frac{H}{M_\mathrm{P}} \right)^4
    e^{5\pi(2m_Q + m_Q^{-1})/2}
    \int_0^{x_\mathrm{max}} \frac{\mathrm{d}y}{y^2}
    (y \cos y - \sin y)
    \left[
        \frac{\partial_y}{m_Q}
        +\frac{m_Q}{y} -1
    \right]
    \int_y^{x_\mathrm{max}} \mathrm{d} \tilde{y} \,
    \nonumber\\
    & \quad \times
    \mathrm{Im} \left[
        W_{\beta, \alpha} (-2i y)^* W_{\beta, \alpha}(-2i \tilde{y})
    \right]
    \left\{
        \left[
             \mathcal{F}(m_Q)
        \right]^3
        \left[
            W_{\beta, \alpha} (-2i\tilde{y})^*
        \right]^3
        +
        \frac{1}{3}
        \left[
             \mathcal{F}(m_Q)
        \right]^2
        \mathcal{F}^*(m_Q)
        W_{\beta, \alpha}(-2i\tilde{y})
        \left[
            W_{\beta, \alpha} (-2i\tilde{y})^*
        \right]^2
    \right\}
    + \mathrm{c.c.}
    \nonumber\\
    & \equiv
    \frac{4 \epsilon_B^2 g^2 C^{(a)}}{k^9}
    \left( \frac{H}{M_\mathrm{P}} \right)^4
    e^{5\pi(2m_Q + m_Q^{-1})/2}
    \left\{
        \left[
             \mathcal{F}(m_Q)
        \right]^3
        \mathcal{N}_{F,\mathrm{eq}}^{(a)}
        +
        \left[
             \mathcal{F}(m_Q)
        \right]^2
        \mathcal{F}^*(m_Q)
        \mathcal{N}_{G,\mathrm{eq}}^{(a)}
        + \mathrm{c.c.}
    \right\}.
\end{align}

\begin{figure}[htpb]
    \includegraphics[clip, width=0.8\columnwidth]{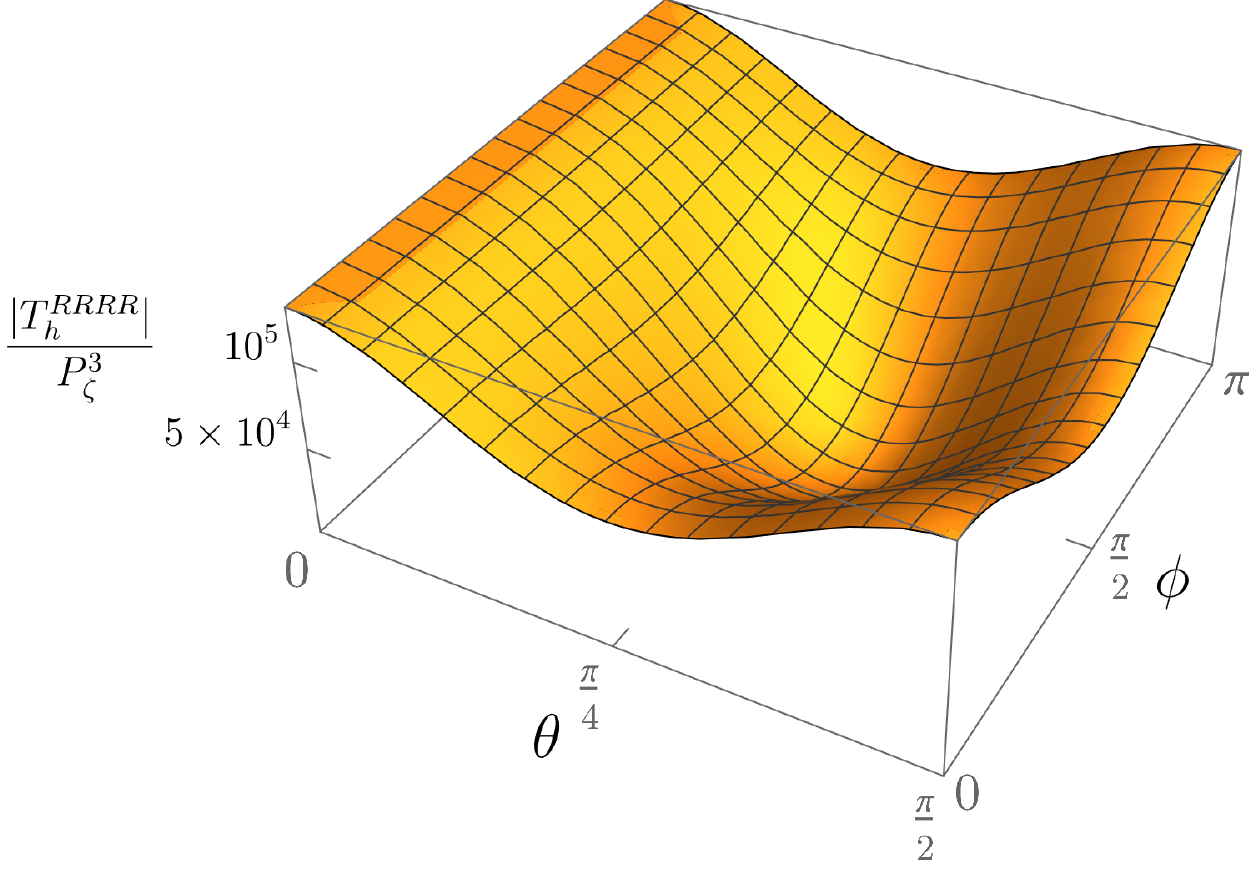}
    \caption{
    Evaluation of $|T_h^{RRRR}|/P_\zeta^3$ as a function of $\theta$ and $\phi$.
    }
    \label{fig: evaluation of Th}
\end{figure}

We evaluate the magnitude of the trispectrum with a set of the parameters
\begin{align}
    g = 10^{-2},
    \quad
    m_Q = 2.8,
    \quad
    r_\mathrm{vac} = 0.01,
    \quad
    \epsilon_B = 6.4 \times 10^{-5},
    \label{eq: test parameters}
\end{align}
where the tensor-to-scalar ratio of the vacuum contribution $r_\mathrm{vac}$ is related to the Hubble rate during inflation $H$ as
$r_\mathrm{vac} = 2H^2/(\pi^2 M_\mathrm{Pl}^2 \mathcal{P}_\zeta)$ 
and we use the scalar power spectrum $\mathcal{P}_\zeta = 2.1 \times 10^{-9}$.
Then, we obtain
\begin{equation}
    \mathcal{F} \simeq -1.9 \times 10^2 + 1.9 \times 10^2 i,
    \quad
    \mathcal{N}_\mathrm{F,eq}^{(a)} 
    \simeq
    1.9 \times 10^{-14} - 1.6 \times 10^{-14} i,
    \quad
    \mathcal{N}_\mathrm{G,eq}^{(a)} 
    \simeq
    -5.7 \times 10^{-15} - 6.0 \times 10^{-15} i,
\end{equation}
and
\begin{equation}
    \frac{T_h^{RRRR}}{P_\zeta^3}
    =
    \frac{k^9 T_h^{RRRR}}{(2\pi^2\mathcal{P}_\zeta)^3}
    \simeq
    6.5 \times 10^4 \, C^{(a)},
    \label{eq: estimation of trispectra (a)}
\end{equation}
which is shown in Fig.~\ref{fig: evaluation of Th} with the dependence of $C^{(a)}$ on $\theta$ and $\phi$.

\subsubsection{non-dynamical components}
\label{subsubsec: evaluation of diagram v non-dynamical}

Next, we consider the contributions from the non-dynamical components in the equilateral limit.
As for the scalar non-dynamical contributions, for example, a part of the trispectra is written as
\begin{align}
    &\frac{\left \langle
        \hat{\psi}_1^R(\tau, \bm{k}_1) \hat{\psi}_1^R(\tau, \bm{k}_2)
        \hat{\psi}_1^R(\tau, \bm{k}_3) \hat{\psi}_{3,t_{3,t_1^3}}^R(\tau, \bm{k}_4)
    \right \rangle_\mathrm{scalar}}{(2 \pi)^3 \delta^{(3)}(\bm{k}_1 + \bm{k}_2 + \bm{k}_3 + \bm{k}_4)}
    \nonumber\\
    &=
    F_\mathrm{s}^{(a)}(\bm{k}_1, \bm{k}_2, \bm{k}_3, \bm{k}_4),
    \nonumber\\
    &=
    -\frac{1}{2}
    \int \mathrm{d}\eta_4 \, G_\psi(\tau, \eta_4, k) \mathcal{D}(\eta_4, k)
    \int \mathrm{d} \tilde{\eta}_4 \, G_t (\eta_4, \tilde{\eta}_4, k)
    \Psi_1^R(\tau, k)^3
    \nonumber\\
    & \quad \times
    \left[
        A_{\mathrm{s},-1,-4} A_{\mathrm{s},-2,-3}
        \left(
            F_{\mathrm{s},1+4}(\tilde{\eta}_4)
            [T_1^{R* \prime}(\tilde{\eta}_4, k)]^2
            T_1^{R*}(\tilde{\eta}_4, k)
            + 
            \left\{
                F_{\mathrm{s},1+4}(\tilde{\eta}_4)
                [T_1^{R*}(\tilde{\eta}_4, k)]^2
                T_1^{R* \prime}(\tilde{\eta}_4, k)
            \right\}'
        \right)
    \right.
    \nonumber\\
    & \qquad
    \left.
        + (\mathrm{permutations\,of\,}\bm{k}_1,\,\bm{k}_2,\,\bm{k}_3)
    \right].
\end{align}
Since $A_{\mathrm{s},a,b}$ is anti-symmetric with respect to $a$ and $b$,
this contribution vanishes by summing up the permutations of the momenta $\bm{k}_1,\,\bm{k}_2,\,\bm{k}_3$.
This is also the case for the other scalar non-dynamical contributions
and then the total contributions from the scalar non-dynamical components vanish in the equilateral limit.

As for the vector non-dynamical contributions, a part of the trispectra is written as
\begin{align}
    &
    \frac{
        \left \langle
            \hat{\psi}_1^R(\tau, \bm{k}_1) \hat{\psi}_1^R(\tau, \bm{k}_2)
            \hat{\psi}_1^R(\tau, \bm{k}_3) \hat{\psi}_{3,t_{3,t_1^3}}^R(\tau, \bm{k}_4)
        \right \rangle_\mathrm{vector}
    }{(2\pi)^3 \delta^{(3)}(\bm{k}_1 + \bm{k}_2 + \bm{k}_3 + \bm{k}_4) }
    \nonumber\\
    &=
    F_\mathrm{v}^{(a)}(\bm{k}_1, \bm{k}_2, \bm{k}_3, \bm{k}_4)
    \nonumber\\
    &=
    -\frac{1}{2}
    \int \mathrm{d}\eta \, G_\psi(\tau, \eta, k) \mathcal{D}(\eta, k)
    \int \mathrm{d} \tilde{\eta} \, G_t (\eta, \tilde{\eta}, k)
    \Psi_1^R(\tau, k)^3
    \sum_{\lambda = R,L}
    \nonumber\\
    & \quad \times
    \left[
        A^\lambda_{\mathrm{v},-4,-1}
        A^\lambda_{\mathrm{v},-2,-3}
        \left(
            F^\lambda_\mathrm{v,1+4}(\tilde{\eta})
            T_1^{R*}(\tilde{\eta},k) 
            \left[ T_1^{R* \prime}(\tilde{\eta},k) \right]^2
        \right.
        +
        \left\{
            F^\lambda_\mathrm{v,1+4}(\tilde{\eta})
            \left[ T_1^{R*}(\tilde{\eta},k) \right]^2
            T_1^{R* \prime}(\tilde{\eta},k)
        \right\}'
    \right)
    \nonumber\\
    & \qquad
    \left.
        + (\mathrm{permutations\,of\,}\bm{k}_1,\,\bm{k}_2,\,\bm{k}_3)
    \right].
\end{align}
Since $A^\lambda_{\mathrm{v},a,b}$ is anti-symmetric with respect to $a$ and $b$,
this contribution also vanishes by summing up the permutations of the momenta $\bm{k}_1,\,\bm{k}_2,\,\bm{k}_3$.
This is also the case for $G_\mathrm{s}^{(a)}(\bm{k}_1, \bm{k}_2, \bm{k}_3, \bm{k}_4)$.

As a result, the contributions from the non-dynamical components all vanish in the equilateral limit.

\subsection{Evaluation of diagram (b)}
\label{subsec: evaluation of diagram b}

Next we evaluate the contribution from diagram (b).
In a similar way to diagram (a), the contributions from the non-dynamical components vanish in diagram (b).
Then, we evaluate the contributions from the dynamical components:
\begin{align}
    &
    \frac{\left \langle
        \hat{\psi}^R(\tau, \bm{k}_1) \hat{\psi}^R(\tau, \bm{k}_2)
        \hat{\psi}^R(\tau, \bm{k}_3) \hat{\psi}^R(\tau, \bm{k}_4)
    \right \rangle
    }
    {(2 \pi)^3 \delta^{(3)}(\bm{k}_1 + \bm{k}_2 + \bm{k}_3 + \bm{k}_4)}
    \nonumber\\
    & \supset
    F_\mathrm{d}^{(b)}(\bm{k}_1, \bm{k}_2, \bm{k}_3, \bm{k}_4)
    \nonumber\\
    & =
    c^{\psi t^3}
    \int \mathrm{d}\eta_4 \, G_\psi(\tau, \eta_4, k)
    \left[ \Psi_1^R(\tau, k) T_1^{R*}(\eta_4, k) \right]^3
    \left[
        I^{\psi t^3}_{-4,-1,-2,-3}
        + (\mathrm{permutations\,of\,}\bm{k}_1,\,\bm{k}_2,\,\bm{k}_3)
    \right].
\end{align}
Here, we parametrize $I^{\psi t^3}$ as
\begin{equation}
    I^{\psi t^3}_{-4,-1,-2,-3}
    + (\mathrm{permutations\,of\,}\bm{k}_1,\,\bm{k}_2,\,\bm{k}_3)
    =
    C^{(b)}_1 + \frac{k \eta}{m_Q} C^{(b)}_2.
\end{equation}
In the equilateral limit, $C^{(b)} _1$ and $C^{(b)}_2$ vanish and then all of the contributions from diagram (b) vanish.

\begin{figure}[htpb]
    \includegraphics[clip, width=0.8\columnwidth]{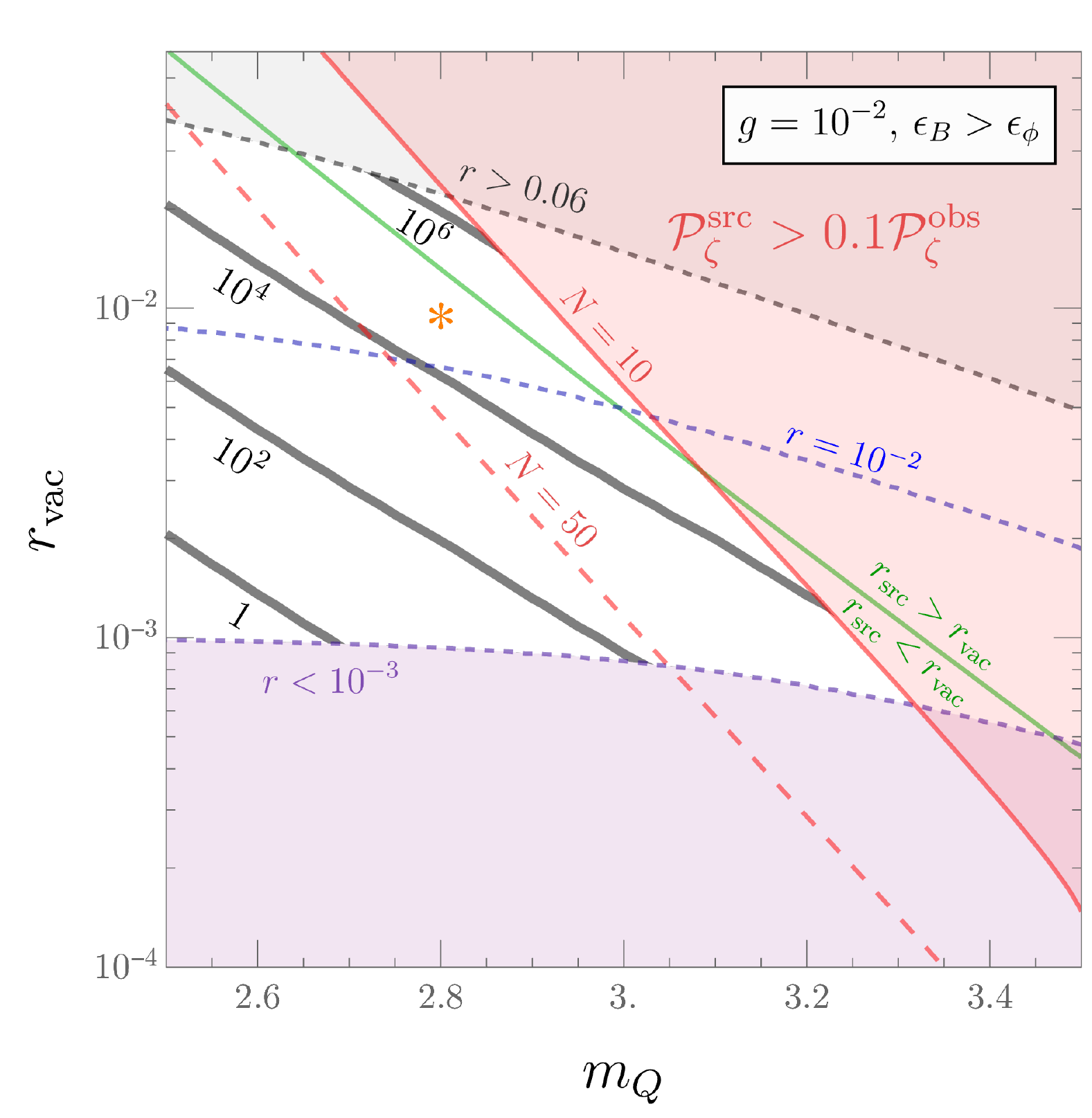}
    \caption{
    Predicted GW trispectrum signals in parameter space constrained so far and searchable in the near future through the CMB power spectrum measurements.
    The Gray lines show the sizes of trispectra without the angular dependence, $T_h^{RRRR}/(P_\zeta^3 C^{(a)})$, in the equilateral limit.
    The light gray region is excluded by the recent Planck/BICEP 2/KECK Array results~\cite{Akrami:2018odb,Ade:2018gkx}, which set the upper bound of the tensor-to-scalar ratio $r < 0.06$.
    In the purple region, $r < 10^{-3}$ is below the sensitivity of the next generation CMB experiments.
    The red region indicates that the sourced curvature perturbations can make non-negligible non-Gaussian contributions to the observed curvature perturbations~\cite{Papageorgiou:2019ecb}, which is limited by the CMB observations~\cite{Ade:2015ava}.
    Above the green line, the power spectrum of the sourced GW dominates that of the vacuum GW.
    The orange star corresponds to the model parameters in Eq.~\eqref{eq: test parameters}.
    }
    \label{fig: Main Result}
\end{figure}

To sum up, in the equilateral limit, the $g_\mathrm{NL}$-type trispectra of the right-handed GW is only contributed by diagram (a).
The magnitude of the trispectra is dependent on the angular configuration of the momenta and the model parameters $g$, $m_Q$, and $r_\mathrm{vac}$.
In Fig.~\ref{fig: Main Result}, we show the constraints on the parameters with $g=0.01$~\cite{Papageorgiou:2019ecb} and the trispectra without the angular dependence $T_h^{RRRR}/(P_\zeta^3 C^{(a)})$ in Eq.~\eqref{eq: estimation of trispectra (a)}.

\begin{figure}[htpb]
    \includegraphics[clip, width=0.8\columnwidth]{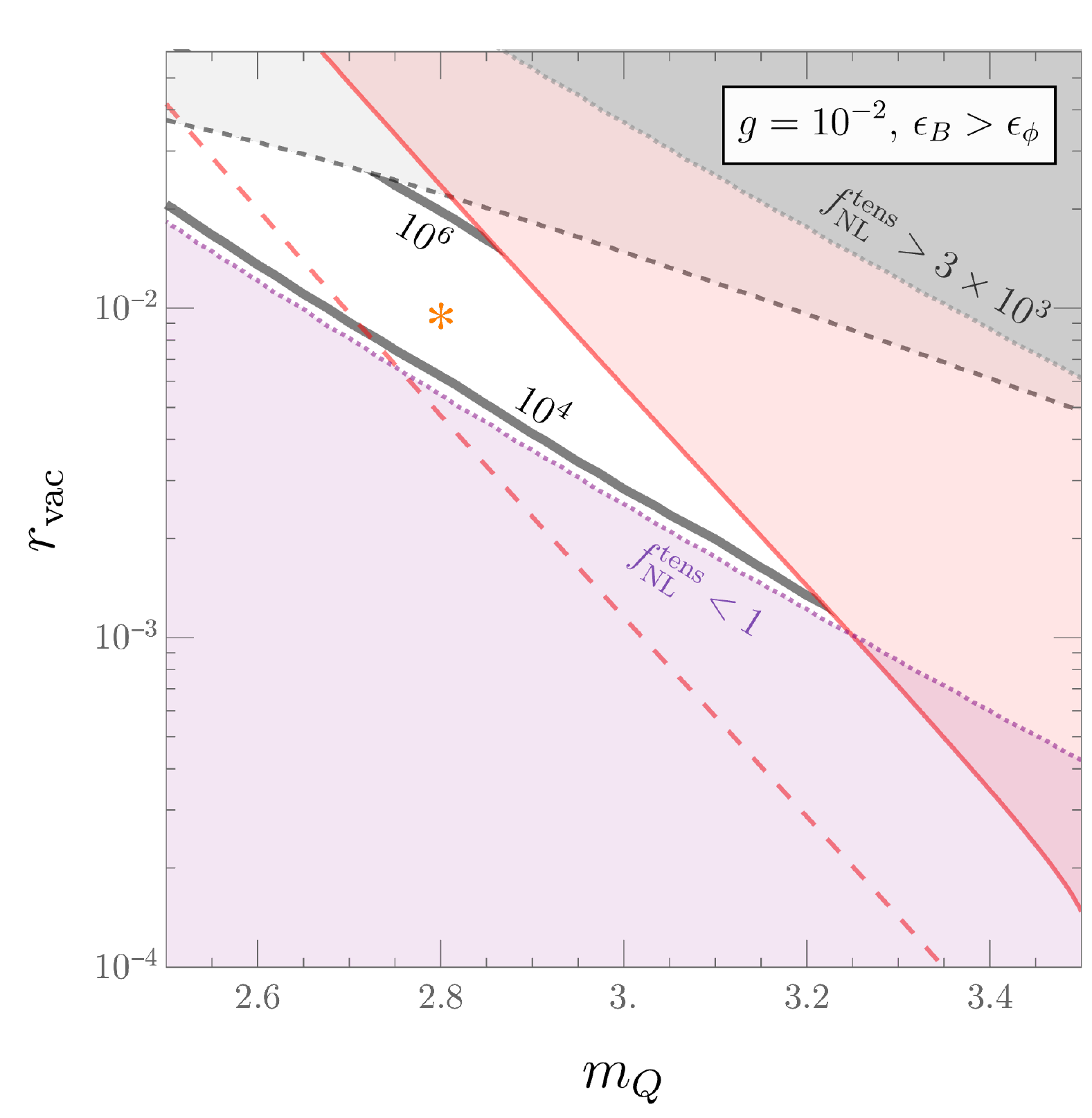}
    \caption{
    Predicted GW trispectrum signals in parameter space constrained so far from the CMB power spectrum and bispectrum and searchable in the near future through the CMB bispectrum measurements.
    The grey region is excluded by the recent Planck upper bound on the tensor nonlinearity parameter \cite{Planck:2019kim}, $f_{\mathrm{NL}}^{\mathrm{tens}} < 3000$.
    The purple region, $f_{\mathrm{NL}}^{\mathrm{tens}} < 1$, is below the sensitivity of the next generation CMB experiments \cite{Shiraishi:2019yux}.
    The other lines are the same as Fig.~\ref{fig: Main Result}. 
    }
    \label{fig: fNL and gNL}
\end{figure}

\section{Summary and discussion}
\label{sec: discussion}

In this paper we have investigated the trispectrum of tensor perturbations sourced by SU(2) gauge fields during inflation.
In particular we focus on four-point vertices of tensor perturbations and $g_\mathrm{NL}$-type trispectrum in the equilateral limit.
Since the four-point vertices come from the self interaction of the SU(2) gauge fields,
this contribution is unique to non-Abelian gauge theory.
Although the non-dynamical components can induce additional four-point vertices after integrating out,
their contribution vanishes in the equilateral limit due to the anti-symmetric form of the interaction.
As a result, we find that $T_h/P_\zeta^3$, which parameterizes the amplitude of the trispectrum, can be as large as $\mathcal{O}(10^6)$ in the parameter regions 
constrained so far and searchable in the near future through the CMB power spectrum and bispectrum measurements
as shown by the black lines in Figs.~\ref{fig: Main Result} and \ref{fig: fNL and gNL}.

With the following rough comparison with the previous scalar-mode detectability analysis, one can see that there is every prospect of detecting such a signal by CMB observations. Now, let us consider the measurements of scalar-mode and tensor-mode primordial trispectra with only large-scale data, $\ell \lesssim 100$, of the CMB temperature field. Then, if the tensor-mode primordial trispectrum is comparable in size to the scalar-mode one, their induced CMB temperature trispectra also have similar magnitudes (because of similar behaviors of scalar-mode and tensor-mode temperature transfer functions at such small $\ell$), yielding similar values of signal to noise ratio. Therefore, from the previous scalar-mode result that $T_\zeta / P_\zeta^3 \sim 10^6$ is detectable by the use of the information with $\ell \lesssim 100$,%
\footnote{This value corresponds to an expected $1\sigma$ error on the original $g_{\rm NL}$, the size parameter of the scalar local-type trispectrum \cite{Bartolo:2015fqz}. Strictly speaking, this trispectrum is amplified at not the equilateral configuration but another configuration that one of four momenta is much smaller than the other three ones; thus, the resulting CMB trispectrum has different shape from our CMB trispectrum. Nevertheless, the impact of the shape difference on the detectability is weak (as we are focusing on a narrow $\ell$ range) and our estimation would be reasonable.}
our target signal, $T_h / P_\zeta^3 \sim 10^6$, is expected to be captured.
Employing B-mode polarization field and high-$\ell$ information can further improve the detectability, achieving more powerful tests in near future.

However, for precise discussions, we, of course, need much more general information of the GW trispectrum including the general configuration of momenta and all combinations of the higher-order perturbations.
In addition, we need to evaluate the effect of the transfer function and derive the CMB signal expected to be observed.
We leave these challenges for future work.

\section*{Acknowledgement}
\label{ack}

We would like to thank Eiichiro Komatsu for useful comments.
This work was supported by the JSPS KAKENHI Grants No. JP18K13537, JP20H05854 (T.\,F.) JP20J20248 (K.\,M.), JP19K14718 (M.\,S.) and JP20H05859 (M.\,S. and I.\,O.). K.\,M. is also supported by World Premier International Research Center Initiative (WPI Initiative), MEXT, Japan and the Program of Excellence in Photon Science. I.\,O acknowledges the support from JSPS Overseas Research Fellowship. M.\,S. acknowledges the Center for Computational Astrophysics, National Astronomical Observatory of Japan, for providing the computing resources of Cray XC50.

\appendix

\section{Polarization tensor}
\label{app: polarization tensor}

For $\bm{k} = k^i = k \delta^{3 i}$,
the left and right-handed polarization vectors are defined by
\begin{equation}
    e_i^{L}(\hat{\bm{k}})
    =
    \frac{1}{\sqrt{2}} \left(
        \begin{array}{c}
            1\\
            i\\
            0
        \end{array}
    \right),
    \quad
    e_i^R(\hat{\bm{k}}) = e_i^{L*}(\hat{\bm{k}}),
\end{equation}

For a general wave number vector:
\begin{equation}
    \bm{k}
    =
    k \left(
        \begin{array}{c}
            \sin \alpha \cos \beta \\
            \sin \alpha \sin \beta \\
            \cos \alpha \\
        \end{array}
    \right),
\end{equation}
$e_i^{L/R}(\hat{\bm{k}})$ are obtained by applying  to $e_{i}^{L/R}(\hat{\bm{z}})$ the following rotation matrix $S(\hat{\bm{k}})$, which transforms $\hat{\bm{z}}$ into $\hat{\bm{k}}$:
\begin{equation}
    S(\hat{\bm{k}})
    =
    \left(
        \begin{array}{ccc}
            \cos \alpha \cos \beta & -\sin \beta & \sin \alpha \cos \beta \\
            \cos \alpha \sin \beta &  \cos \beta & \sin \alpha \sin \beta \\
           -\sin \alpha              &             0 & \cos \alpha              \\
        \end{array}
    \right).
\end{equation}
Note that
\begin{equation}
    S(-\hat{\bm{k}})
    =
    \left(
        \begin{array}{ccc}
            \cos (\pi - \alpha) \cos (\beta + \pi) &
            -\sin (\beta + \pi) &
            \sin (\pi - \alpha) \cos (\beta + \pi) 
            \\
            \cos (\pi - \alpha) \sin (\beta + \pi) &
            \cos (\beta + \pi) &
            \sin (\pi - \alpha) \sin (\beta + \pi)
            \\
            -\sin (\pi - \alpha) &
            0 &
            \cos (\pi - \alpha)                     
            \\
        \end{array}
    \right)
    =
    \left(
        \begin{array}{ccc}
            \cos \alpha \cos \beta &  \sin \beta & -\sin \alpha \cos \beta \\
            \cos \alpha \sin \beta & -\cos \beta & -\sin \alpha \sin \beta \\
           -\sin \alpha              &             0 & -\cos \alpha              \\
        \end{array}
    \right),
\end{equation}
and then
\begin{equation}
    e_i^{R/L} (-\hat{\bm{k}})
    =
    S_{i j}(-\hat{\bm{k}}) e_j^{R/L} (\hat{\bm{z}})
    =
    S_{i j}(\hat{\bm{k}}) e_j^{L/R} (\hat{\bm{z}})
    =
    e_i^{L/R} (\hat{\bm{k}})
    =
    e_i^{R/L*} (\hat{\bm{k}}).
\end{equation}

The polarization tensors are constructed from the polarization vectors as
\begin{equation}
    e_{i j}^{R/L}(\hat{\bm{k}})
    =
    e_{i}^{R/L}(\hat{\bm{k}}) e_{j}^{R/L}(\hat{\bm{k}}).
\end{equation}

These polarization vectors and tensors satisfy

\begin{align}
    i \epsilon^{i k l} k^l e_k^{R/L}(\bm{k}) 
    &=
    \pm k e_i^{R/L}(\bm{k}),
    \\
    i \epsilon^{i k l} k^l e_{j k}^{R/L}(\bm{k}) 
    &=
    \pm k e_{i j}^{R/L}(\bm{k}).
\end{align}

Next, in order to investigate the trispectra, we consider four momenta $\bm{k}_1$, $\bm{k}_2$, $\bm{k}_3$, and $\bm{k}_4$ satisfying $\bm{k}_1 + \bm{k}_2 + \bm{k}_3 + \bm{k}_4 = \bm{0}$.
Note that we cannot generally choose the coordinate system where all of the momenta lie in the $x$-$y$ plane.

\begin{figure}[htpb]
    \includegraphics[clip, width=0.45\columnwidth]{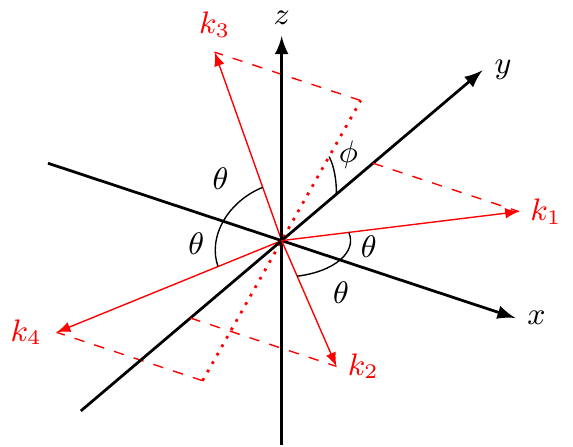}
\caption{
The momentum configuration in the equilateral limit.
}
\label{fig: equilateral}
\end{figure}

In the following, we consider the equilateral limit for simplicity and use the parameterization as shown in Fig.~\ref{fig: equilateral}:
\begin{align}
    \bm{k}_1 
    &=
    k (\cos \theta, \, \sin \theta, 0),
    \\
    \bm{k}_2 
    &=
    k (\cos \theta, \, -\sin \theta, 0),
    \\
    \bm{k}_3 
    &=
    k (-\cos \theta , \, \sin \theta \cos \phi, \, \sin \theta \sin \phi), 
    \\
    \bm{k}_4
    &=
    k (-\cos \theta , \, -\sin \theta \cos \phi, \, -\sin \theta \sin \phi).
\end{align}
In the equilateral limit, we always can take this coordinate system by setting $\hat{\bm{x}} \parallel \bm{k}_1 + \bm{k}_2$ and $\hat{\bm{z}} \perp \bm{k}_1$.
Without loss of generality, we can set $0 < \theta < \pi/2$ and $0 \leq \phi < \pi$ by exchanging $\bm{k}_i$.
Note that this coordinate system avoids $\bm{k}_i \parallel \pm \hat{\bm{z}}$, where the phases of polarization vector and tensor are not well-defined.

For example, the products of the polarization tensors have complex phases in general:
\begin{align}
    e_{i j}^R(\hat{\bm{k}}_1) e_{j k}^R(\hat{\bm{k}}_2) 
    e_{k l}^R(\hat{\bm{k}}_3) e_{l i}^R(\hat{\bm{k}}_4)
    &=
    \frac{\sin^4 \theta (\cos \theta \sin (\phi/2) + i \cos (\phi/2))^4
    (\cos \phi + i \cos \theta \sin \phi)^4}
    {( \cos^2 \theta + \cos^2 \phi \sin^2 \theta)^2},
    \\
    e_{i j}^R(\hat{\bm{k}}_1) e_{i j}^R(\hat{\bm{k}}_2) 
    e_{k l}^R(\hat{\bm{k}}_3) e_{k l}^R(\hat{\bm{k}}_4)
    &=
    \frac{\sin^8 \theta (\cos \phi + i \cos \theta \sin \phi)^4}
    {( \cos^2 \theta + \cos^2 \phi \sin^2 \theta)^2}
    .
\end{align}

\section{Fourth order Lagrangian from non-dynamical variables}
\label{App: 4th Lagrangian from non-dynamical}

In this appendix, we consider the contribution to the fourth order Lagrangian from the non-dynamical variable $\delta A_0^a \equiv a^{-1}[\partial_a Y + Y_a]$.
We denote the action including $\delta A_0^a$ as
\begin{equation}
    S_\mathrm{nd}
    =
    \int \mathrm{d}\tau \mathrm{d}^3x \,
    L_\mathrm{nd}.
\end{equation}
In order to evaluate the relevant fourth order Lagrangian, we have to expand $L_\mathrm{nd}$ up to $\mathcal{O}(Y^2, Y \psi t, Y t^2)$.

First, $\mathcal{O}(Y^2)$ term $L_{\mathrm{nd},Y^2}$ comes from $FF$ term and can be written  as
\begin{align}
    L_{\mathrm{nd},Y^2}
    =&
    \frac{1}{2} Y \partial^4 Y
    - g^2 a^2 Q^2 Y \partial^2 Y
    \nonumber\\
    &
    -\frac{1}{2} Y_a \partial^2 Y_a
    + g a Q \epsilon^{a b c} (\partial_c Y_a) Y_b
    + g^2 a^2 Q^2 Y_a Y_a.
\end{align}

$\mathcal{O}(Y \psi t, Y t^2)$ terms come from the gauge kinetic term $FF$:
\begin{equation}
    L_{\mathrm{nd},FF}
    =
    g \epsilon^{a b c} Y \partial_b ( t'_{a i} t_{c i})
    - g \epsilon^{a b c} Y_b t'_{a i} t_{c i}
    +\frac{2m_Q H}{M_\mathrm{P}} \psi_{i j} ( Y_a + \partial_a Y)
    \left[
        \frac{\tau}{m_Q} \partial_i t_{a j}' 
        -\epsilon^{i a b} \left(t_{b j}' + \frac{t_{b j}}{\tau}\right)
    \right].
\end{equation}
Since $F \tilde{F}$ is a total derivative, the part of the Chern-Simons term including $\delta A_0^a$ has the form of $\chi_0 \partial_i K_i$, where $K_i$ is a function of the gauge field.
Then, this contribution vanishes after an integration by parts.

To summarize above, the relevant Lagrangian including the scalar mode $Y$ is
\begin{align}
    L_\mathrm{nd,s}
    =&
    \frac{1}{2} Y \partial^4 Y
    - \frac{m_Q^2}{\tau^2} Y \partial^2 Y
    + g \epsilon^{a b c} Y \partial_b ( t'_{a i} t_{c i})
    +\frac{2m_Q H}{M_\mathrm{P}} \psi_{i j}\partial_a Y
    \left[
        \frac{\tau}{m_Q} \partial_i t_{a j}' 
        -
        \epsilon^{i a b} \left(t_{b j}' + \frac{t_{b j}}{\tau}\right)
    \right],
\end{align}
and that including the vector mode $Y_a$ is
\begin{align}
    L_\mathrm{nd,v}
    =&
    -\frac{1}{2} Y_a \partial^2 Y_a
    - \frac{m_Q}{\tau} \epsilon^{a b c} (\partial_c Y_a) Y_b
    + \frac{m_Q^2}{\tau^2} Y_a Y_a
    - g \epsilon^{a b c} Y_b t'_{a i} t_{c i}
    \nonumber\\
    &+\frac{2m_Q H}{M_\mathrm{P}} \psi_{i j} Y_a
    \left[
        \frac{\tau}{m_Q} \partial_i t_{a j}' 
        -
        \epsilon^{i a b} \left(t_{b j}' + \frac{t_{b j}}{\tau}\right)
    \right].
\end{align}

Next, we obtain the fourth order Lagrangian originating from $Y$ and $Y_a$ by completing the square in the Fourier space.
We decompose $Y$ and $Y_a$ as
\begin{align}
    Y(\tau, \bm{x})
    &=
    \int \frac{\mathrm{d}^3k}{(2\pi)^3} Y_{\bm{k}} (\tau) e^{i \bm{k} \cdot \bm{x}},
    \\
    Y_a(\tau, \bm{x}) 
    &=
    \sum_{\lambda = L, R} \int \frac{\mathrm{d}^3k}{(2\pi)^3} Y_{\lambda,\bm{k}}(\tau) e^\lambda_a (\hat{\bm{k}}) e^{i \bm{k} \cdot \bm{x}},
\end{align}
where $e^\lambda_a(\bm{k})$ is a polarization vector shown in App.~\ref{app: polarization tensor}.

For the scalar mode, we can rewrite $S_\mathrm{nd,s} \equiv \int \mathrm{d}\tau \mathrm{d}^3x \, L_\mathrm{nd,s}$ as
\begin{align}
    S_\mathrm{nd,s}
    =
    \int \frac{\mathrm{d}\tau \mathrm{d}^3 k}{(2\pi)^3}
    &\left[ 
        Y_{-\bm{k}} \left( \frac{k^4}{2} + \frac{m_Q^2 k^2}{\tau^2} \right) Y_{\bm{k}}
    \right.
    \nonumber\\
    & +
        \int \frac{\mathrm{d}^3 p}{(2\pi)^3} Y_{\bm{k}}
        \left\{
            c^{t^3} t_{-\bm{p}}' t_{-\bm{k}+\bm{p}}
            (p - |\bm{p} - \bm{k}|) 
            e_{i j}^R(-\hat{\bm{p}}) e_{i j}^R(\widehat{\bm{p} - \bm{k}})
        \right.
    \nonumber\\
    &   
    \left.
        \left.
            \qquad \qquad \quad \ 
            + c^{\psi t^2} \psi_{-\bm{k}+\bm{p}}
            \left(
                \frac{\tau}{m_Q} k^j k^k t_{-\bm{p}}'
                e_{i j}^R(-\hat{\bm{p}}) e_{i k}^R(\widehat{\bm{p}-\bm{k}})
                -
                (p - |\bm{p} - \bm{k}|)
                \left( t_{-\bm{p}}' + \frac{t_{-\bm{p}}}{\tau} \right)
                e_{i j}^R(-\hat{\bm{p}}) e_{i j}^R(\widehat{\bm{p}-\bm{k}})
            \right)
        \right\}
    \right].
\end{align}

By completing the square and integrating out $Y$,
we obtain the fourth order Lagrangian coming from the non-dynamical scalar variables as
\begin{align}
    S_\mathrm{nd,s}
    =&
    -\int \frac{\mathrm{d}\tau \mathrm{d}^3 k \mathrm{d}^3 p \mathrm{d}^3 q \mathrm{d}^3 r}{4(2\pi)^9}
    \delta^{(3)}( \bm{k} + \bm{p} + \bm{q} + \bm{r})
    \left( 
        \frac{|\bm{k} + \bm{p}|^4}{2} 
        + \frac{m_Q^2 |\bm{k} + \bm{p}|^2}{\tau^2} 
    \right)^{-1}
    \nonumber\\
    & \times \left\{
            c^{t^3} t_{\bm{p}}' t_{\bm{k}}
            (p - k) 
            e_{i j}^R(\hat{\bm{p}}) e_{i j}^R(\hat{\bm{k}})
    \right.
    \nonumber\\
    & \quad
    \left.
        \left.
            + c^{\psi t^2} \psi_{\bm{k}}
            \left(
                \frac{\tau}{m_Q} (k^j + p^j) (k^k + p^k) t_{\bm{p}}'
                e_{i j}^R(\hat{\bm{p}}) e_{i k}^R(\hat{\bm{k}})
                -
                (p - k) \left(t_{\bm{p}}' + \frac{t_{\bm{p}}}{\tau}\right)
                e_{i j}^R(\hat{\bm{p}}) e_{i j}^R(\hat{\bm{k}})
            \right)
        \right\}
    \right.
    \nonumber\\
    & \times \left\{
            c^{t^3} t_{\bm{q}}' t_{\bm{r}}
            (q - r) 
            e_{l m}^R(\hat{\bm{q}}) e_{l m}^R(\hat{\bm{r}})
    \right.
    \nonumber\\
    & \quad 
    \left.
        \left.
            + c^{\psi t^2} \psi_{\bm{r}}
            \left(
                \frac{\tau}{m_Q} (q^m + r^m) (q^n + r^n) t_{\bm{q}}'
                e_{l m}^R(\hat{\bm{q}}) e_{l n}^R(\hat{\bm{r}})
                -
                (q - r) \left(t_{\bm{q}}' + \frac{t_{\bm{q}}}{\tau}\right)
                e_{l m}^R(\hat{\bm{q}}) e_{l m}^R(\hat{\bm{r}})
            \right)
        \right\}
    \right. .
\end{align}
Therefore, $\mathcal{O}(\psi t^3)$ and $\mathcal{O}(t^4)$ terms are
\begin{align}
    S_{\mathrm{nd, s}}^{\psi t^3}
    =
    -\int 
    &
    \frac{\mathrm{d}\tau \mathrm{d}^3 k \mathrm{d}^3 p \mathrm{d}^3 q \mathrm{d}^3 r}{2(2\pi)^9}
    \delta^{(3)}( \bm{k} + \bm{p} + \bm{q} + \bm{r})
    \left( 
        \frac{|\bm{k} + \bm{p}|^4}{2} 
        + \frac{m_Q^2 |\bm{k} + \bm{p}|^2}{\tau^2} 
    \right)^{-1}
    \nonumber\\
    &   
    \times c^{\psi t^2} \psi_{\bm{k}}
    \left(
        \frac{\tau}{m_Q} (k^j + p^j) (k^k + p^k) t_{\bm{p}}'
        e_{i j}^R(\hat{\bm{p}}) e_{i k}^R(\hat{\bm{k}})
        -
        (p - k) \left(t_{\bm{p}}' + \frac{t_{\bm{p}}}{\tau}\right)
        e_{i j}^R(\hat{\bm{p}}) e_{i j}^R(\hat{\bm{k}})
    \right)
    \nonumber\\
    &
    \times c^{t^3} t_{\bm{q}}' t_{\bm{r}}
    (q - r) e_{l m}^R(\hat{\bm{q}}) e_{l m}^R(\hat{\bm{r}})
    \nonumber\\
    =
    -\int 
    &
    \frac{\mathrm{d}\tau \mathrm{d}^3 k \mathrm{d}^3 p \mathrm{d}^3 q \mathrm{d}^3 r}{2(2\pi)^9}
    \delta^{(3)}( \bm{k} + \bm{p} + \bm{q} + \bm{r})
    F_\mathrm{s}(\tau, |\bm{k} + \bm{p}|)
    \nonumber\\
    &   
    \times \psi_{\bm{k}}
    \left(
        \frac{\tau}{m_Q} t_{\bm{p}}'
        S_\mathrm{s}(\bm{p},\bm{k})
        -
        \frac{c^{\psi t^2}}{c^{t^3}}
        \left(t_{\bm{p}}' + \frac{t_{\bm{p}}}{\tau}\right)
        A_\mathrm{s}(\bm{p},\bm{k})
    \right)
    t_{\bm{q}}' t_{\bm{r}}
    A_\mathrm{s}(\bm{q}, \bm{r}),
    \\
    S_{\mathrm{nd, s}}^{t^4}
    =
    -\int 
    &
    \frac{\mathrm{d}\tau \mathrm{d}^3 k \mathrm{d}^3 p \mathrm{d}^3 q \mathrm{d}^3 r}{4(2\pi)^9}
    \delta^{(3)}( \bm{k} + \bm{p} + \bm{q} + \bm{r})
    \left( 
        \frac{|\bm{k} + \bm{p}|^4}{2} 
        + \frac{m_Q^2 |\bm{k} + \bm{p}|^2}{\tau^2} 
    \right)^{-1}
    \nonumber\\
    &
    \times c^{t^3} t_{\bm{p}}' t_{\bm{k}}
    (p - k) e_{i j}^R(\hat{\bm{p}}) e_{i j}^R(\hat{\bm{k}})
    c^{t^3} t_{\bm{q}}' t_{\bm{r}}
    (q - r) e_{l m}^R(\hat{\bm{q}}) e_{l m}^R(\hat{\bm{r}})
    \nonumber\\
    =
    -\int
    &
    \frac{\mathrm{d}\tau \mathrm{d}^3 k \mathrm{d}^3 p \mathrm{d}^3 q \mathrm{d}^3 r}{4(2\pi)^9}
    \delta^{(3)}( \bm{k} + \bm{p} + \bm{q} + \bm{r})
    F_\mathrm{s}(\tau, |\bm{k}+ \bm{p}|)
    t_{\bm{p}}' t_{\bm{k}}
    t_{\bm{q}}' t_{\bm{r}}
    A_\mathrm{s}(\bm{p}, \bm{k})
    A_\mathrm{s}(\bm{q}, \bm{r}),
\end{align}
where
\begin{align}
    F_\mathrm{s}(\tau, |\bm{k}+ \bm{p}|) 
    & \equiv
    \left( 
        \frac{|\bm{k} + \bm{p}|^4}{2} 
        + \frac{m_Q^2 |\bm{k} + \bm{p}|^2}{\tau^2} 
    \right)^{-1},
    \\
    S_\mathrm{s}(\bm{p}, \bm{k})
    & \equiv
    c^{\psi t^2} (k^j + p^j) (k^k + p^k) 
    e_{i j}^R(\hat{\bm{p}}) e_{i k}^R({\hat{\bm{k}}}).
    \\
    A_\mathrm{s}(\bm{p}, \bm{k})
    & \equiv
    c^{t^3}(p - k) 
    e_{i j}^R(\hat{\bm{p}}) e_{i j}^R(\hat{\bm{k}}).
\end{align}

For the vector mode, we can rewrite $S_{\mathrm{nd, v}} \equiv \int \mathrm{d}\tau \mathrm{d}^3 x\,L_{\mathrm{nd, v}}$ as
\begin{align}
    S_{\mathrm{nd, v}}
    =
    \int \frac{\mathrm{d}\tau \mathrm{d}^3 k}{(2\pi)^3}
    \sum_{\lambda = R, L}
    &\left[
        Y_{\lambda, -\bm{k}}
        \left(
            \frac{k^2}{2} + s_\lambda k \frac{m_Q}{\tau} + \frac{m_Q^2}{\tau^2}
        \right)
        Y_{\lambda, \bm{k}}
    \right.
    \nonumber\\
    &+ \int \frac{\mathrm{d}^3 p}{(2\pi)^3} 
    Y_{\lambda, \bm{k}} e_a^\lambda (\hat{\bm{k}})
    \left\{
        c^{t^3} \epsilon^{a b c} 
        e_{b i}^R(\widehat{-\bm{k}-\bm{p}}) e_{c i}^R(\hat{\bm{p}})
        t_{\bm{p}} t'_{-\bm{k}-\bm{p}}
    \right.
    \nonumber\\
    & \qquad  \qquad \qquad \qquad \quad
    \left.
        \left.
            + c^{\psi t^2} \psi_{-\bm{k}-\bm{p}} e_{i j}^R(\widehat{-\bm{k}-\bm{p}})
            \left(
                i p^i \frac{\tau}{m_Q} e_{a j}^R(\hat{\bm{p}}) t'_{\bm{p}}
                -
                \epsilon^{i a b} 
                \left[ t'_{\bm{p}} + \frac{t_{\bm{p}}}{\tau}\right]
                e_{b j}^R(\hat{\bm{p}})
            \right)
        \right\}
    \right],
\end{align}
where $s_{R/L} = \pm 1$.

By completing the square and integrating out $Y_a$, we obtain the fourth order Lagrangian coming from the non-dynamical vector variables as
\begin{align}
    S_{\mathrm{nd, v}}
    =
    - \int 
    \frac{\mathrm{d} \tau \mathrm{d}^3 k \mathrm{d}^3 p \mathrm{d}^3 q \mathrm{d}^3 r}
    {4 (2\pi)^9}
    &\delta^{(3)}(\bm{k} + \bm{p} + \bm{q} + \bm{r})
    \sum_{\lambda = R, L}
    \left(
        \frac{|\bm{k}+\bm{p}|^2}{2} + s_\lambda |\bm{k}+\bm{p}| \frac{m_Q}{\tau} + \frac{m_Q^2}{\tau^2}
    \right)^{-1}
    \nonumber\\
    \times e_a^\lambda(\widehat{-\bm{k}-\bm{p}})
    &\left\{
        c^{t^3} \epsilon^{a b c} 
        e_{b d}^R(\hat{\bm{k}}) e_{c d}^R(\hat{\bm{p}})
        t_{\bm{p}} t'_{\bm{k}}
    \right.
    \nonumber\\
    &
    \left.
        + c^{\psi t^2} \psi_{\bm{k}} e_{c d}^R(\hat{\bm{k}})
        \left(
            i p^c \frac{\tau}{m_Q} e_{a d}^R(\hat{\bm{p}}) t'_{\bm{p}}
            -
            \epsilon^{a b c} 
            \left[ t'_{\bm{p}} + \frac{t_{\bm{p}}}{\tau}\right]
            e_{b d}^R(\hat{\bm{p}})
        \right)
    \right\}
    \nonumber\\
    \times e_i^\lambda(\widehat{\bm{k}+\bm{p}})
    &\left\{
        c^{t^3} \epsilon^{i j k} 
        e_{j l}^R(\hat{\bm{r}}) e_{k l}^R(\hat{\bm{q}})
        t_{\bm{q}} t'_{\bm{r}}
    \right.
    \nonumber\\
    &
    \left.
        + c^{\psi t^2} \psi_{\bm{r}} e_{k l}^R(\hat{\bm{r}})
        \left(
            i q^k \frac{\tau}{m_Q} e_{i l}^R(\hat{\bm{q}}) t'_{\bm{q}}
            -
            \epsilon^{i j k} 
            \left[ t'_{\bm{q}} + \frac{t_{\bm{q}}}{\tau}\right]
            e_{j l}^R(\hat{\bm{q}})
        \right)
    \right\}.
\end{align}
Therefore, $\mathcal{O}(\psi t^3)$ and $\mathcal{O}(t^4)$ terms are
\begin{align}
    S_{\mathrm{nd, v}}^{\psi t^3}
    =
    - \int 
    &
    \frac{\mathrm{d} \tau \mathrm{d}^3 k \mathrm{d}^3 p \mathrm{d}^3 q \mathrm{d}^3 r}
    {2 (2\pi)^9}
    \delta^{(3)}(\bm{k} + \bm{p} + \bm{q} + \bm{r})
    \sum_{\lambda = R, L}
    \left(
        \frac{|\bm{k}+\bm{p}|^2}{2} + s_\lambda |\bm{k}+\bm{p}| \frac{m_Q}{\tau} + \frac{m_Q^2}{\tau^2}
    \right)^{-1}
    \nonumber\\
    & \times
    e_a^\lambda(\widehat{-\bm{k}-\bm{p}})
    c^{\psi t^2} \psi_{\bm{k}} e_{c d}^R(\hat{\bm{k}})
    \left(
        i p^c \frac{\tau}{m_Q} e_{a d}^R(\hat{\bm{p}}) t'_{\bm{p}}
        -
        \epsilon^{a b c} 
        \left[ t'_{\bm{p}} + \frac{t_{\bm{p}}}{\tau}\right]
        e_{b d}^R(\hat{\bm{p}})
    \right)
    \nonumber\\
    & \times
    e_i^\lambda(\widehat{\bm{k}+\bm{p}})
    c^{t^3} \epsilon^{i j k} 
    e_{j l}^R(\hat{\bm{r}}) e_{k l}^R(\hat{\bm{q}})
    t_{\bm{q}} t'_{\bm{r}}
    \nonumber\\
    =
    - \int 
    &
    \frac{\mathrm{d} \tau \mathrm{d}^3 k \mathrm{d}^3 p \mathrm{d}^3 q \mathrm{d}^3 r}
    {2 (2\pi)^9}
    \delta^{(3)}(\bm{k} + \bm{p} + \bm{q} + \bm{r})
    \sum_{\lambda = R, L}
    F_\mathrm{v}^\lambda(\tau, |\bm{k}+\bm{p}|)
    \nonumber\\
    & \times
    \psi_{\bm{k}}
    \left(
        \frac{\tau}{m_Q} P_\mathrm{v}^\lambda(\bm{p},\bm{k}) t'_{\bm{p}}
        -
        \frac{c^{\psi t^2}}{c^{t^3}}
        A_\mathrm{v}^\lambda(\bm{p},\bm{k})
        \left[ t'_{\bm{p}} + \frac{t_{\bm{p}}}{\tau}\right]
    \right)
    A_\mathrm{v}^\lambda(\bm{q},\bm{r})
    t_{\bm{q}} t'_{\bm{r}},
    \\
    S_{\mathrm{nd, v}}^{t^4}
    =
    - \int 
    &
    \frac{\mathrm{d} \tau \mathrm{d}^3 k \mathrm{d}^3 p \mathrm{d}^3 q \mathrm{d}^3 r}
    {4 (2\pi)^9}
    \delta^{(3)}(\bm{k} + \bm{p} + \bm{q} + \bm{r})
    \sum_{\lambda = R, L}
    \left(
        \frac{|\bm{k}+\bm{p}|^2}{2} + s_\lambda |\bm{k}+\bm{p}| \frac{m_Q}{\tau} + \frac{m_Q^2}{\tau^2}
    \right)^{-1}
    \nonumber\\
    & \times e_a^\lambda(\widehat{-\bm{k}-\bm{p}})
    c^{t^3} \epsilon^{a b c} 
    e_{b d}^R(\hat{\bm{k}}) e_{c d}^R(\hat{\bm{p}})
    t_{\bm{p}} t'_{\bm{k}}
    e_i^\lambda(\widehat{\bm{k}+\bm{p}})
    c^{t^3} \epsilon^{i j k} 
    e_{j l}^R(\hat{\bm{r}}) e_{k l}^R(\hat{\bm{q}})
    t_{\bm{q}} t'_{\bm{r}}
    \nonumber\\
    =
    -\int 
    &
    \frac{\mathrm{d}\tau \mathrm{d}^3 k \mathrm{d}^3 p \mathrm{d}^3 q \mathrm{d}^3 r}{4(2\pi)^9}
    \delta^{(3)}( \bm{k} + \bm{p} + \bm{q} + \bm{r})
    \sum_{\lambda = R, L}
    F^\lambda_\mathrm{v}(\tau, |\bm{k}+ \bm{p}|)
    A^\lambda_\mathrm{v}(\bm{p}, \bm{k}) t_{\bm{p}} t_{\bm{k}}'
    A^\lambda_\mathrm{v}(\bm{q}, \bm{r}) t_{\bm{q}} t_{\bm{r}}',
\end{align}
where
\begin{align}
    F^\lambda_\mathrm{v}(\tau, |\bm{k}+ \bm{p}|) 
    & \equiv 
    \left(
        \frac{|\bm{k}+\bm{p}|^2}{2} + s_\lambda |\bm{k}+\bm{p}| \frac{m_Q}{\tau} + \frac{m_Q^2}{\tau^2}
    \right)^{-1},
    \\
    P^\lambda_\mathrm{v}(\bm{p}, \bm{k})
    & \equiv
    e_a^\lambda(\widehat{-\bm{k}-\bm{p}})
    c^{\psi t^2} i p^b
    e_{b c}^R(\hat{\bm{k}}) e_{a c}^R(\hat{\bm{p}}),
    \\
    A^\lambda_\mathrm{v}(\bm{p}, \bm{k})
    & \equiv
    e_a^\lambda(\widehat{-\bm{k}-\bm{p}})
    c^{t^3} \epsilon^{a b c} 
    e_{b d}^R(\hat{\bm{k}}) e_{c d}^R(\hat{\bm{p}}).
\end{align}

\small
\bibliographystyle{apsrev4-2}
\bibliography{Ref}

\end{document}